\documentclass[11pt,a4paper]{article}%
\usepackage{amssymb,amsmath, amsfonts}
\usepackage{graphicx,graphics}
\usepackage{mathtools}
\usepackage{bbold}
\usepackage[english]{babel}
\usepackage[utf8]{inputenc}
\usepackage{epsfig,url}
\usepackage{bbm,theorem}
\usepackage{a4wide}
\usepackage{color}
\usepackage{enumerate}
\usepackage{calrsfs}
\usepackage[bookmarks=true]{hyperref}
\usepackage{bookmark}
\usepackage{amsmath}
\usepackage{amsfonts}
\usepackage{amssymb}
\usepackage{verbatim}
\usepackage{graphicx}%
\providecommand{\U}[1]{\protect\rule{.1in}{.1in}}
\DeclareMathAlphabet{\pazocal}{OMS}{zplm}{m}{n}
\newtheorem{theorem}{Theorem}[section]

{\theorembodyfont{\upshape}
\newtheorem{remark}[theorem]{Remark}

}
\numberwithin{equation}{section}
\numberwithin{theorem}{section}
\newcommand{\qed}{\hfill$\Box$}

\newcommand{\R}{{\mathbb R}}

\newcommand{\ep}{\varepsilon}

\DeclareMathOperator{\Tr}{Tr}

\newcommand{\eps}{{\varepsilon}}

\newcommand{\beq}{\begin{equation}}
\newcommand{\eeq}{\end{equation}}
\newcommand{\beqs}{\begin{eqnarray}}
\newcommand{\eeqs}{\end{eqnarray}}

\newcounter{jlisti}

\begin{document}

\title{Long time asymptotics for homoenergetic solutions of the Boltzmann equation. Hyperbolic-dominated case.}
\author{Richard D. James \thanks{E-mail: \texttt{james@umn.edu}}, Alessia Nota \thanks{E-mail:
\texttt{nota@iam.uni-bonn.de}} , Juan J. L. Vel\'azquez \thanks{E-mail:
\texttt{velazquez@iam.uni-bonn.de}}\\
$\,^{\ddag}$\emph{Department of Aerospace Engineering and Mechanics} \\University of Minnesota, \\\emph{University of Bonn, Institute for Applied Mathematics}\\Endenicher Allee 60, D-53115 Bonn, Germany}
\date{\today }
\maketitle

\begin{abstract}
In this paper we continue the formal analysis of the long-time asymptotics of 
the homoenergetic solutions for the Boltzmann equation that we began in \cite{JNV2}. 
They have the form $f\left(  x,v,t\right)  =g\left(v-L\left(  t\right)  x,t\right)  $ where $L\left(  t\right)  =A\left(I+tA\right)  ^{-1}$ where $A$ is a constant matrix.   
Homoenergetic solutions satisfy an integro-differential equation which contains, in addition to the classical Boltzmann collision operator, a linear hyperbolic term.  
Depending on the properties of the collision kernel the collision and the hyperbolic terms might be of the same order of magnitude as $t\to\infty$, or the collision term could be the dominant one for large times, or the hyperbolic term could be the largest.
 The first case has been  rigorously studied in  \cite{JNV1}. Formal asymptotic expansions in the second case have been obtained in \cite{JNV2}. All the solutions obtained in this case can be approximated by Maxwellian distributions with changing temperature.  
 
In this paper we focus in the case where the hyperbolic terms are much larger than the collision term for large times (hyperbolic-dominated behavior). In the hyperbolic-dominated case it does not seem to be possible to describe in a simple way all the long time asymptotics of the solutions, but we discuss several physical situations and formulate precise conjectures. 
We give explicit formulas for the relationship between density, temperature and entropy for these solutions. 
These formulas differ greatly from the  ones at equilibrium.
\end{abstract}
\tableofcontents

\bigskip

\bigskip

\bigskip

\bigskip
\section{Introduction \label{sect1}}

In this paper we continue the study of the long time asymptotics of the homoenergetic solutions of 
the Boltzmann equation which do not exhibit self-similar behaviours.  
We began this analysis in \cite{JNV1} where the self-similar case was considered and, in \cite{JNV2}, where solutions behaving asymptotically as Maxwellian distributions with changing temperature were studied. 

The class of solutions under consideration is motivated by
an invariant manifold of solutions of the equations of classical molecular
dynamics with certain symmetry properties (\cite{md, viscometry}).

We shortly recall  the main properties of this manifold (we refer to \cite{JNV1} for a more detailed description). Suppose that we consider a matrix $A\in M_{3\times3}\left(  \mathbb{R}%
\right)$, satisfying $\det(I+tA)>0$
for $t\in\lbrack0,a)$ with $a>0$, and the orthonormal vectors $e_{1},e_{2},e_{3}$ in
$\mathbb{R}^{3}$. We consider $M$ \textit{simulated atoms} with positive masses $m_{1},\dots,m_{M}$ 
subject to the equations of molecular dynamics yielding solutions
$y_{k}(t)\in\mathbb{R}^{3},\ 0\leq t<a,\ k=1,\dots,M$. 
Moreover, we denote as $y_{\nu,k}(t)$, $\nu=(\nu_{1},\nu_{2},\nu_{3})\in
\mathbb{Z}^{3}$ the positions of the \textit{non-simulated atoms} which are given by 
\begin{equation}
y_{\nu,k}(t)=y_{k}(t)+(I+tA)(\nu_{1}e_{1}+\nu_{2}e_{2}+\nu_{3}e_{3}),\quad
\nu=(\nu_{1},\nu_{2},\nu_{3})\in\mathbb{Z}^{3},\ k=1,\dots,M. \label{nonsim}%
\end{equation}
For $k=1,\dots,M$ we denote as $f_{k}:\cdots\mathbb{R}^{3}\times\mathbb{R}%
^{3}\times{\mathbb{R}^{3}}\cdots\rightarrow\mathbb{R}$ the force on simulated
atom $k$ which depends on the positions of all the
atoms. Assuming that $f_{k}$ satisfies the usual conditions of
frame-indifference and permutation invariance \cite{md} we have 
\begin{align}
m_{k}\ddot{y}_{k}  &  =f_{k}(\dots,y_{\nu_{1},1},\dots,y_{\nu_{1},M}%
,\dots,y_{\nu_{2},1},\dots,y_{\nu_{2},M},\dots),\label{md_sim}\\
{y}_{k}(0)  &  ={y}_{k}^{0},\quad\dot{y}_{k}(0)={v}_{k}^{0},\quad
k=1,\dots,M.\nonumber
\end{align}
Using \eqref{nonsim} we can reduce \eqref{md_sim} to a system of ODEs for the motions of
the simulated atoms, i.e. $y_{k}(t),$ $k=1,\dots,M$. 

It is shown in \cite{md} and \cite{viscometry}
that in spite of the fact that the motions of the nonsimulated atoms are only
given by the formulas (\ref{nonsim}), the equations of molecular dynamics
(\ref{md_sim}) are exactly satisfied for each nonsimulated atom. 

\begin{figure}[th]
\label{fig1}\centering
\includegraphics [scale=0.3]{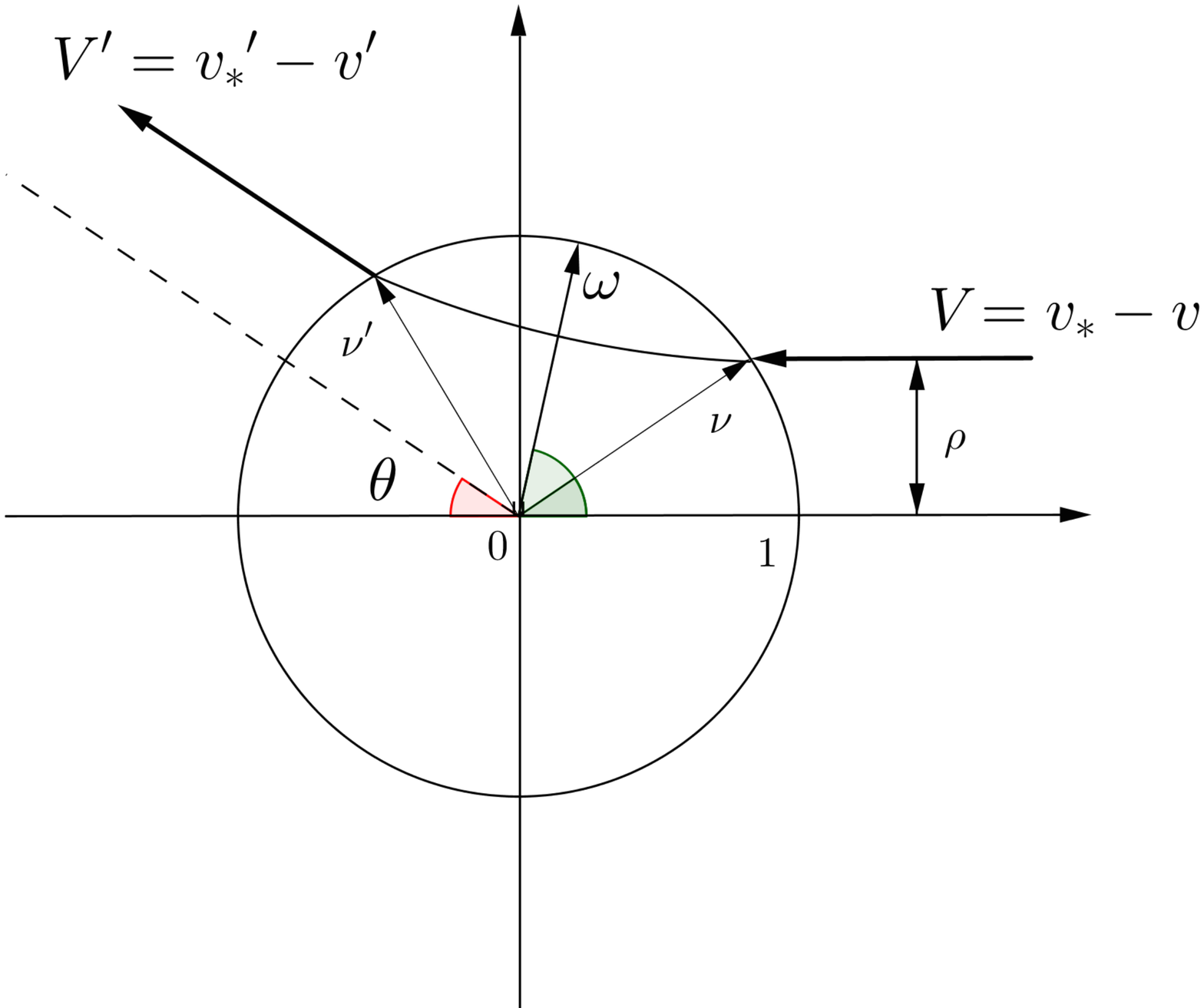}\caption{The two-body scattering. 
 The scalar $\rho\in[-1,1]$ is the impact parameter, 
 and $\theta=\theta(\rho, |V|)$ is the
scattering angle. The scattering vector of (\ref{CM1}), (\ref{CM2}) is the
unit vector $\omega=\omega(v,V)$.} %
\end{figure}

 As discussed in \cite{JNV1}, these results on molecular dynamics have a simple
counterpart in terms of the molecular density function of the kinetic
theory. The classical Boltzmann equation reads as
\begin{align}
\partial_{t}f+v\partial_{x}f  &  =\mathbb{C}f\left(  v\right)
\ \ ,\ \ f=f\left(  t,x,v\right) \nonumber\\
\mathbb{C}f\left(  v\right)   &  =\int_{\mathbb{R}^{3}}dv_{\ast}\int_{S^{2}%
}d\omega B\left(  n\cdot\omega,\left\vert v-v_{\ast}\right\vert \right)
\left[  f^{\prime}f_{\ast}^{\prime}-f_{\ast}f\right] , \ \label{A0_0}%
\end{align}
where $S^{2}$ is the unit sphere in $\mathbb{R}^{3}$ and $n=n\left(
v,v_{\ast}\right)  =\frac{\left(  v-v_{\ast}\right)  }{\left\vert v-v_{\ast
}\right\vert }.$ Let be $(v,v_{\ast})$ a pair of incoming velocities (see Figure \ref{fig1}). The outgoing velocities $(v^{\prime},v_{\ast
}^{\prime})$ are given by the collision rule
\begin{align}
v^{\prime}  &  =v+\left(  \left(  v_{\ast}-v\right)  \cdot\omega\right)
\omega,\label{CM1}\\
v_{\ast}^{\prime}  &  =v_{\ast}-\left(  \left(  v_{\ast}-v\right)  \cdot
\omega\right)  \omega, \label{CM2}%
\end{align}
where  $\omega=\omega(v,V)$ is the unit vector bisecting the angle between the incoming
relative velocity $V=v_{\ast}-v$ and the outgoing relative velocity
$V^{\prime}=v_{\ast}^{\prime}-v^{\prime}$ as specified in Figure \ref{fig1}. 

We will use in the rest of the paper the standard convention $f=f\left( t, x, v\right)  ,\ f_{\ast}=f\left(  t,
x, v_{\ast}\right)  ,\ f^{\prime}=f\left(  t, x, v^{\prime}\right)
,\ \ f_{\ast}^{\prime}=f\left(  t, x, v_{\ast}^{\prime}\right)  $.

The collision kernel $B\left(  n\cdot\omega,\left\vert v-v_{\ast}\right\vert
\right)  $ is proportional to the cross section for the scattering problem
associated to the collision between two particles. 
We will assume that it is homogeneous in 
$\left\vert v-v_{\ast}\right\vert $ and we will denote its homogeneity by
$\gamma,\ $i.e.,
\begin{equation}
B\left(  n  \cdot\omega,\lambda\left\vert v-v_{\ast}\right\vert \right)  =\lambda
^{\gamma}B\left( n \cdot\omega,\left\vert v-v_{\ast}\right\vert \right)
,\ \ \lambda>0. \label{S8E7}%
\end{equation}

The homogeneity $\gamma$ is related to the properties of the interaction potential between particles. 
We recall that in the standard literature in kinetic theory (cf.~\cite{V02}), interaction
potentials with the form $V\left(  x\right)  =\frac{1}{\left\vert x\right\vert
^{\nu-1}}$ have homogeneity $\gamma=\frac{\nu-5}{\nu-1}$ for the kernel $B$.

It is possible to find solutions of the Boltzmann equation \eqref{A0_0}
that have the same statistics as the molecular dynamics simulation for discrete systems described above (see \eqref{nonsim}). We refer to \cite{JNV1} for details.  
Thus, the analogous of the ansatz (\ref{nonsim}) in terms of the
particle velocities 
can be written as: 
\begin{equation}
f(t,x,v)=g(t,v-A(I+tA)^{-1}x).\label{ansatz}%
\end{equation}
The term $A(I+tA)^{-1}$ arises from conversion to the Eulerian form of the
kinetic theory.

An alternative way of deriving \eqref{ansatz} is by means of the theory of \textit{equidispersive
solutions} for the Boltzmann equation.  
These are solutions of the Boltzmann equation with the form%
\begin{equation}
f\left(  t,x,v\right)  =g\left(  t,w\right)  \text{ \ \ with }w=v-\xi\left(
t,x\right)  .\ \label{HomSol}%
\end{equation}

Under mild smoothness conditions, solutions with the form
(\ref{HomSol}) exist if $\xi(t,x)=A(I+tA)^{-1}x$ (cf.~\cite{JNV1}). Formally, if $f$
is a solution of the Boltzmann equation (\ref{A0_0}) of the form
(\ref{ansatz}) the function $g$ satisfies
\begin{equation}
\partial_{t}g-\big(L\left(  t\right)  w\big)\cdot\partial_{w}g=\mathbb{C}g\left(
w\right)  \label{D1_0}%
\end{equation}
where the collision operator $\mathbb{C}$ is defined as in \eqref{A0_0}. These
solutions are called \textit{homoenergetic solutions} and were introduced by
Galkin \cite{Galkin1} and Truesdell \cite{T} and later considered in 
\cite{Bobylev75}, \cite{Bobylev76}, \cite{BobCarSp}, \cite{CercArchive},
\cite{Cerc2000}, \cite{Cerc2002}, \cite{Galkin1}, \cite{Galkin2}, \cite{Galkin3}, \cite{garzo},
\cite{Nikol1}, \cite{Nikol2}, \cite{T}, \cite{TM}.
\bigskip

The properties of the solutions of (\ref{D1_0}) for large times $t$ depend
greatly on the homogeneity of the kernel yielding the cross section of the
collision operator $\mathbb{C}g.$ In \cite{JNV1} we have focused on
the analysis of solutions of (\ref{D1_0}) for which the terms $L\left(
t\right)  w\cdot\partial_{w}g$ and $\mathbb{C}g\left(  w\right)$ are of the same order of magnitude.   We have rigorously proved in \cite{JNV1} the existence of self-similar solutions in the class of homoenergetic flows when 
the collision kernel describes the interaction between Maxwell molecules, which corresponds to homogeneity $\gamma=0$. In
all the cases when such self-similar solutions exist, the terms $L\left(
t\right)  w\cdot\partial_{w}g$ and $\mathbb{C}g\left(  w\right)  $ have a
comparable size as $t\rightarrow\infty.$

In \cite{JNV2}  we considered the case in which the collision terms are the dominant ones as $t\to\infty$.
The solutions obtained in that paper are approximately Maxwellians, with a time dependent temperature whose evolution is obtained using
a suitable adaptation of the standard Hilbert expansion.

\medskip
Finally, it turns out that there are also choices of $L\left(  t\right)$ and collision kernels $B$ for
which the scaling properties of the different terms imply that the hyperbolic
terms are much more important than the collision terms.  This hyperbolic-dominated case is the one studied in this paper. We emphasize that in this case there is a large variety of different asymptotic behaviours for the homoenergetic flows.

For instance, in some
particular cases discussed in this paper the effect of the collision
terms becomes negligible as $t\rightarrow\infty$ and the distribution of
particle velocities is asymptotically given by the hyperbolic terms in
(\ref{D1_0}), namely $L\left(  t\right)  w\cdot\partial_{w}g$. However we will
find also situations in which, in spite of becoming increasingly small as
$t\rightarrow\infty$, the collision term $\mathbb{C}g\left(  w\right)  $ plays
a crucial role characterizing the distribution of velocities of the particles. This is due to the fact that the mean free path increases to $\infty$ as $t\to \infty$ but, at the same time, for each given particle the probability of having infinitely many collisions in the time interval $(0,\infty)$ is equal to one.  
This case is very interesting, 
but it is also the case for which we have more fragmentary results (cf.~Section \ref{sec:hypcases}). In order to shed some light on the behaviour of the particle distribution $g$,
we introduce in Section \ref{sec:domhyper} a simplified model, which does not correspond to
any Boltzmann equation, but contains several of the main characteristics which
can be found for homoenergetic solutions in the case in which the hyperbolic
terms are dominant. The behavior of the velocity distributions that
we obtain in that case is quite different from the ones obtained in the
previous cases, since these distributions cannot be approximated by
Maxwellians, but they are also not self-similar. 
\bigskip

The plan of the paper is the following. In Section \ref{hom} 
we summarize the most relevant properties of homoenergetic solutions of the Boltzmann
equation which have been obtained in \cite{JNV1, JNV2}. 

 In Section \ref{sec:hypcases} we
discuss several results that we have obtained about hyperbolic-dominated
homoenergetic flows. In particular, in Subsection \ref{sec:domhyper}, we describe some results for a simple toy
model for which the long time asymptotics is hyperbolic-dominated. It is
possible to obtain analytically information about the long time asymptotics of
the solutions of the toy model and hopefully this could shed some light about
the behaviour of more complex hyperbolic-dominated fluxes. Section \ref{sec:entropy} contains
formulas about the behaviour of the entropy for the solutions derived here and in \cite{JNV1, JNV2}. 
This allows to estimate how far from equilibrium are the asymptotics 
of the obtained solutions. 

In Section \ref{sec:tableconcl} we conclude presenting an overview of the results obtained in this paper as well as in \cite{JNV1, JNV2}.


\section{Homoenergetic solutions of the Boltzmann equation}
\label{hom}

We assume that the molecular density function $f\left(  t,x,v\right)  $ satisfies \eqref{A0_0}, namely
\begin{align*}
\partial_{t}f+v\partial_{x}f  &  =\mathbb{C}f\left(  v\right)
\ \ ,\ \ f=f\left(  t,x,v\right) \\\mathbb{C
}f\left(  v\right)   &  =\int_{\mathbb{R}^{3}}dv_{\ast}\int_{S^{2}}d\omega
B\left(  n\cdot\omega,\left\vert v-v_{\ast}\right\vert \right)  \left[
f^{\prime}f_{\ast}^{\prime}-f_{\ast}f\right] . \
\end{align*}

Homoenergetic solutions of (\ref{A0_0})  defined in \cite{Galkin1} and
\cite{T} (cf. also \cite{TM}) are solutions of the Boltzmann equation having
the form%
\begin{equation}
f\left(  t,x,v\right)  =g\left(  t,w\right)  \text{ \ \ with }w=v-\xi\left(
t,x\right) . \label{B1_0}%
\end{equation}

In order to have solutions of (\ref{A0_0})
with the form (\ref{B1_0}) for a sufficiently large class of initial data we
must have%
\begin{equation}
\frac{\partial\xi_{k}}{\partial x_{j}}\text{ independent on }x\text{ and
}\partial_{t}\xi+\xi\cdot\nabla\xi=0 . \label{B2_0}%
\end{equation}

The first condition implies that $\xi$ is an affine function on $x$. In \cite{JNV1}, \cite{JNV2}
we restricted to the case in which $\xi$ is a
linear function of $x$ for simplicity. The general solution of this equation is  
\begin{equation}
\xi\left(  t,x\right)  =L\left(  t\right)  x+B(t), \ \label{B4_0}%
\end{equation}
where $L\left(  t\right)  \in M_{3\times3}\left(  \mathbb{R}\right)  $ is a
$3\times3$ real matrix and $B(t) \in \R^3$. Then, the second equation in (\ref{B2_0}) holds if and only if
\begin{equation}
\frac{dL\left(  t\right)  }{dt}+\left(  L\left(  t\right)  \right)  ^{2}=0,
\quad L (0) = A, \label{B3_0}%
\end{equation}
for some initial condition $A\in M_{3\times3}\left(  \mathbb{R}\right) $, and 
\begin{equation}
\frac{dB\left(  t\right)  }{dt}+L\left(  t\right)B\left(  t\right) =0,
\quad B(0) = B_0, \label{eq:B}%
\end{equation}
for some initial condition $B_0\in \R^3$.

The unique continuous solutions of
\eqref{B3_0}-\eqref{eq:B} are given by
\begin{align}
&L\left(  t\right)  =\left(  I+tA\right)  ^{-1}A=A\left(  I+tA\right)
^{-1},\ \label{B7_0}\\&
 B\left(  t\right)  =\left(  I+tA\right)  ^{-1}B_0=B_0\left(  I+tA\right)
^{-1}\label{eq:solB}
\end{align}
defined on a maximal interval of existence $[0,a)$. On the interval $[0,a)$,
$\det\left(  I+tA \right) >0 $.

We observe that the function $g(t,w)$ solves \eqref{D1_0} even if we choose $\xi(x,t)$ as the affine function \eqref{B4_0}.

\bigskip

We now recall the classification of homoenergetic flows which has
been obtained in \cite{JNV1} (cf. Theorem 3.1). More precisely, we describe
the long time asymptotics of the linear term of $\xi\left(  t,x\right)$, namely $L\left(  t\right)
x=\left(  I+tA\right)  ^{-1}Ax$ (cf. (\ref{B4_0}) and (\ref{B7_0})). Notice that the linear part is the only one which plays a relevant role in the long time  asymptotics of the solution $g(t,w)$.

\medskip

It has been proved in \cite{JNV1} (cf. Theorem 3.1), using all the possible Jordan decompositions, that  for  any matrix $A \in M_{3 \times
3}(\mathbb{R}) $ such that $\det(I + t A) >0$ for $t\ge0$ the possible long time asymptotics of the matrix $L(t) = (I +
tA)^{-1}A$ is one of the following: 
\medskip

\noindent Case (i) Homogeneous dilatation:
\begin{equation}
L(t) = \frac{1}{t} I + O\bigg( \frac{1}{t^{2}} \bigg) \quad\mathrm{as}\ t
\to\infty.\label{T1E1}%
\end{equation}
\noindent Case (ii) Cylindrical dilatation (K=0), or Case (iii) Cylindrical
dilatation and shear ($K \ne0$):
\begin{equation}
L(t) = \frac{1}{t} \left(
\begin{array}
[c]{ccc}%
1 & 0 & K\\
0 & 1 & 0\\
0 & 0 & 0
\end{array}
\right)  + O\bigg( \frac{1}{t^{2}} \bigg) \quad\mathrm{as}\ t \to
\infty.\label{T1E2}%
\end{equation}
\noindent Case (iv). Planar shear:
\begin{equation}
L(t) = \frac{1}{t} \left(
\begin{array}
[c]{ccc}%
0 & 0 & 0\\
0 & 0 & K\\
0 & 0 & 1
\end{array}
\right)  + O\bigg( \frac{1}{t^{2}} \bigg) \quad\mathrm{as}\ t \to
\infty.\label{T1E3}%
\end{equation}
\noindent Case (v). Simple shear:
\begin{equation}
L(t) = \left(
\begin{array}
[c]{ccc}%
0 & K & 0\\
0 & 0 & 0\\
0 & 0 & 0
\end{array}
\right)  , \quad K \ne0.\label{T1E5}%
\end{equation}
\noindent Case (vi). Simple shear with decaying planar dilatation/shear:
\begin{equation}
L(t) = \left(
\begin{array}
[c]{ccc}%
0 & K_{2} & 0\\
0 & 0 & 0\\
0 & 0 & 0
\end{array}
\right)  + \frac{1}{t} \left(
\begin{array}
[c]{ccc}%
0 & K_{1} K_{3} & K_{1}\\
0 & 0 & 0\\
0 & K_{3} & 1
\end{array}
\right)  + O\bigg( \frac{1}{t^{2}} \bigg) , \quad K_{2} \ne0.\label{T1E6}%
\end{equation}
\noindent Case (vii). Combined orthogonal shear:
\begin{equation}
L(t) = \left(
\begin{array}
[c]{ccc}%
0 & K_{3} & K_{2} - t K_{1} K_{3}\\
0 & 0 & K_{1}\\
0 & 0 & 0
\end{array}
\right) , \quad K_{1} K_{3} \ne0.\label{T1E4}%
\end{equation}
\medskip

\begin{remark}
As we observed in \cite{JNV1} there are 
choices of $A\in M_{3\times3}\left(  \mathbb{R}\right)  $ for which $L\left(
t\right)  $ blows up in finite time, but we will restrict here to the case $\det(I + t A) >0$ for all $t \ge0$. 
\end{remark}

\subsection{ Hydrodynamical fields for homoenergetic solutions}
We introduce the following quantities. The density $\rho$
 \begin{equation}\label{eq:density}
 \rho(t,x)=\int_{\R^3} f(t,x,v) dv,
 \end{equation}
 the average velocity $V$ at each point $x$ and time $t$ by means of 
\begin{equation}\label{eq:averagevel}
 \rho\left(  t,x\right)  V\left(  t,x\right)  =\int_{\mathbb{R}^{3}
}f\left(  t,x,v\right)  vdv.
\end{equation}
and the internal 
energy $\varepsilon$ (or temperature) at each point $x$ and time $t$ by means of 
\begin{equation}\label{eq:intenergy}
\rho\left(  t,x\right)  \varepsilon\left(  t,x\right)  =\int_{\mathbb{R}^{3}%
}f\left(  t,x,v\right)  \left(  v-V\left(  t,x\right)  \right)  ^{2}dv.
\end{equation}
We will denote as $c:=v-V$ the random or peculiar velocity, namely the deviation of the velocity of a single particle from the average velocity. 
We define the stress tensor and the  heat flux in terms of $c$ and $f$ as
\begin{equation}\label{eq:momflow2}
\quad M_{ij}=\int_{\R^3}  c_ic_j f(t,x,v) \,dv, \qquad \quad i,j=1,2,3
\end{equation}
\begin{equation}
q_i(t,x)= \int_{\R^3}\!c_i|c|^2f(t,x,v)\,dv \qquad \quad i=1,2,3 \label{eq:heatflux}.
\end{equation}

The Boltzmann equation \eqref{A0_0} implies the following not closed system of 5 scalar conservation laws (mass, momentum and kinetic energy). See for instance \cite{CIP}, \cite{TM}.
\begin{align}\label{eq:conslaw1}
&\frac{\partial\rho}{\partial t} +\sum_{j=1}^{3}\frac{\partial}{\partial x_j}(\rho V_j)=0; \\&
\frac{\partial}{\partial t}(\rho V_i) +\sum_{j=1}^{3}\frac{\partial}{\partial x_j}(\rho V_iV_j+M_{ij})=0 \quad i=1,2,3 \label{eq:conslaw2} \\&
\frac{\partial}{\partial t}(\rho \eps) +\sum_{j=1}^{3}\frac{\partial}{\partial x_j}\big(\rho\eps V_j +q_j\big)+\sum_{i=1}^{3}\sum_{j=1}^{3} M_{ij}\frac{\partial V_i}{\partial x_j}=0 \label{eq:conslaw3}.
\end{align}

We now rewrite the  quantities defined above in the case of homoenergetic flows (cf.~\eqref{B1_0}).  
 We can assume without loss of generality (see Remark \ref{rk:zeroaverage} below) that the class of homoenergetic solutions $g(t,w)$ we consider satisfies $\int_{\R^3}g(t,w)wdw=0$. Then, we have
 \begin{equation}\label{eq:homdensity}
 \rho(t)=\int_{\R^3} g(t,w) dw, 
 \end{equation}
 \begin{equation}\label{eq:homaveragevel}
V\left(  t,x\right)  =\xi(x,t),
\end{equation}
\begin{equation}\label{eq:homintenergy}
\rho\left(  t\right)  \varepsilon\left(  t\right)  =\frac 1 2 \int_{\mathbb{R}^{3}%
}g\left(  t,w\right)  \vert w\vert  ^{2}dw,
\end{equation}
and 
\begin{align}
&M_{ij}(t)=\int_{\R^3} \!\! w_i w_j g\,dw \qquad\quad i,j=1,2,3 \label{eq:hommomflow}\\ &
q_i(t)= \frac 1 2\int_{\R^3}\!w_i|w|^2g\,dw \qquad \quad i=1,2,3. \label{eq:homheatflux}
\end{align}
Therefore, the conservation laws \eqref{eq:conslaw1}-\eqref{eq:conslaw3} become
\begin{align}\label{eq:homconslaw1}
&\frac{\partial}{\partial t}\rho(t) +\sum_{j=1}^{3}\frac{\partial}{\partial x_j}(\rho \xi_j)=0; \\&
\frac{\partial}{\partial t}(\rho(t) \xi_i) +\sum_{j=1}^{3}\frac{\partial}{\partial x_j}(\rho \xi_i\xi_j)=0 \quad i=1,2,3 \label{eq:homconslaw2} \\&
\frac{\partial}{\partial t}(\rho(t) \eps) +\sum_{j=1}^{3}\frac{\partial}{\partial x_j}\big(\rho\eps \xi_j \big)+\sum_{i=1}^{3}\sum_{j=1}^{3} M_{ij}\frac{\partial \xi_i}{\partial x_j}=0 \label{eq:homconslaw3}.
\end{align}
We notice that the equation for the average velocity does not depend on the stress tensor and the equation for the internal energy does not contain the heat flux. 

Using now \eqref{B4_0}, i.e. $\xi\left(  t,x\right)  =L\left(  t\right)  x+B(t)$, in \eqref{eq:homconslaw1}-\eqref{eq:homconslaw3} we obtain
\begin{align}
\label{eq:homconslaw1_bis}
&\frac{\partial}{\partial t}\rho(t) +\rho(t) \Tr(L(t))=0; \\& 
\rho(t)\left(\frac{\partial \xi_{i}(t)}{\partial t}+\sum_{j=1}^{3}\xi_{j}\frac{\partial \xi_{i}(t)}{\partial x_j}\right)=0 \quad i=1,2,3 \label{eq:homconslaw2_bis} \\&
\rho(t) \frac{\partial \eps(t)}{\partial t} +\sum_{i=1}^{3}\sum_{j=1}^{3} M_{ij}(t)L_{ij}(t) = \rho(t) \frac{\partial \eps(t)}{\partial t} +\Tr((ML)(t))=0 \label{eq:homconslaw3_bis},
\end{align}
where $\Tr(ML)=\int_{\R^3} w \cdot Lw \, g\,dw$. We observe that \eqref{eq:homconslaw1_bis} gives the evolution of the density in time, \eqref{eq:homconslaw2_bis} holds due to \eqref{B2_0} and \eqref{eq:homconslaw3_bis} yields the evolution of the internal energy. 

Notice that \eqref{eq:homconslaw1_bis} implies 
\begin{equation}
\rho\left(  t\right)  =\rho\left(  0\right)  \exp\left(  -\int_{0}%
^{t}\Tr \left(  L\left(  s\right)  \right)  ds\right)  . \label{S8E6}%
\end{equation}
which gives the evolution of the density in terms of the properties of the flow described by $L(t)$.  On the contrary, we cannot compute the evolution of the 
internal energy $\varepsilon\left(  t\right)$ without obtaining before some additional information concerning the velocity distribution function $g(t,w)$. 
\medskip

\begin{remark}\label{rk:zeroaverage}
Multiplying \eqref{D1_0} by $w_k$, for $k=1,2,3$ and integrating with respect to $w$ we obtain 
\begin{equation*}
\frac{\partial}{\partial t}\big(\int_{\R^3}g w_k dw \big)+L_{k,j}\int_{\R^3}g w_j dw+\Tr(L)\int_{\R^3}g w_k dw=0,
\end{equation*}
We can assume without loss of generality that $\int_{\R^3}g_0 w_k dw=0$ at $t=0$ for $k=1,2,3$ changing, if needed, the value of $B_0$. This implies $\int_{\R^3}g w_k dw=0$ for any $t>0$. \end{remark}

\bigskip


\section{Homoenergetic flows with hyperbolic-dominated behavior for large
$t\rightarrow\infty$}
\label{sec:hypcases}

\bigskip

For some homoenergetic flows satisfying  \eqref{D1_0} and some choices of the homogeneity $\gamma$ 
of the collision kernel $B$ we can expect the hyperbolic term $-L\left(  t\right)  w\cdot\partial
_{w}g$ to be much larger than the collision term $\mathbb{C}g\left(
t,w\right)  $ (cf.~\eqref{D1_0}). The information that we have about those
homoenergetic flows is much more fragmentary than the one 
 obtained in the cases in which the collision terms are the dominant
ones for large $t$ (cf.~\cite{JNV2}) or in the case in which the hyperbolic and collision terms
have the same order of magnitude as $t\rightarrow\infty$ (cf.~\cite{JNV1}).  In the first case we
have obtained that the asymptotics of the velocity distributions of the
 homoenergetic flows are given by Maxwellian distributions with
time dependent temperatures, whose evolution in time is computed using
suitable adaptations of the classical Hilbert expansions. In the second case, we have proved in \cite{JNV1} the existence of  non Maxwellian self-similar solutions which describe the long time asymptotics of the particle distributions. 
Nevertheless, in both
cases, there is a time dependent characteristic velocity $\left\vert
w\right\vert \approx\ell\left(  t\right)  $ which characterizes the scale of
velocities in which most of the particles of the system as well as the energy
is contained at a given time. More precisely, this means that if we denote by
$M(t),\, E(t)$ the mass and the energy of the system respectively, we have
\[
M(t)\simeq\int_{\delta \ell(t)\leq|w|\leq\frac{1}{\delta} \ell(t)}g(t,w)dw
\]
and
\[
E(t)\simeq\int_{\delta \ell(t)\leq|w|\leq\frac{1}{\delta} \ell(t)}|w|^{2} g(t,w)dw
\]
for some $\delta>0$ small enough.

In several of the cases dominated by the hyperbolic terms discussed in this
section we will argue that such scale $\ell\left(  t\right)  $ containing most
of the particles and the energy does not exist, or there is no single time-dependent velocity scale that characterizes both mass and energy as $t\rightarrow\infty.$

\bigskip

\begin{figure}[th]
\centering
\includegraphics [scale=0.5]{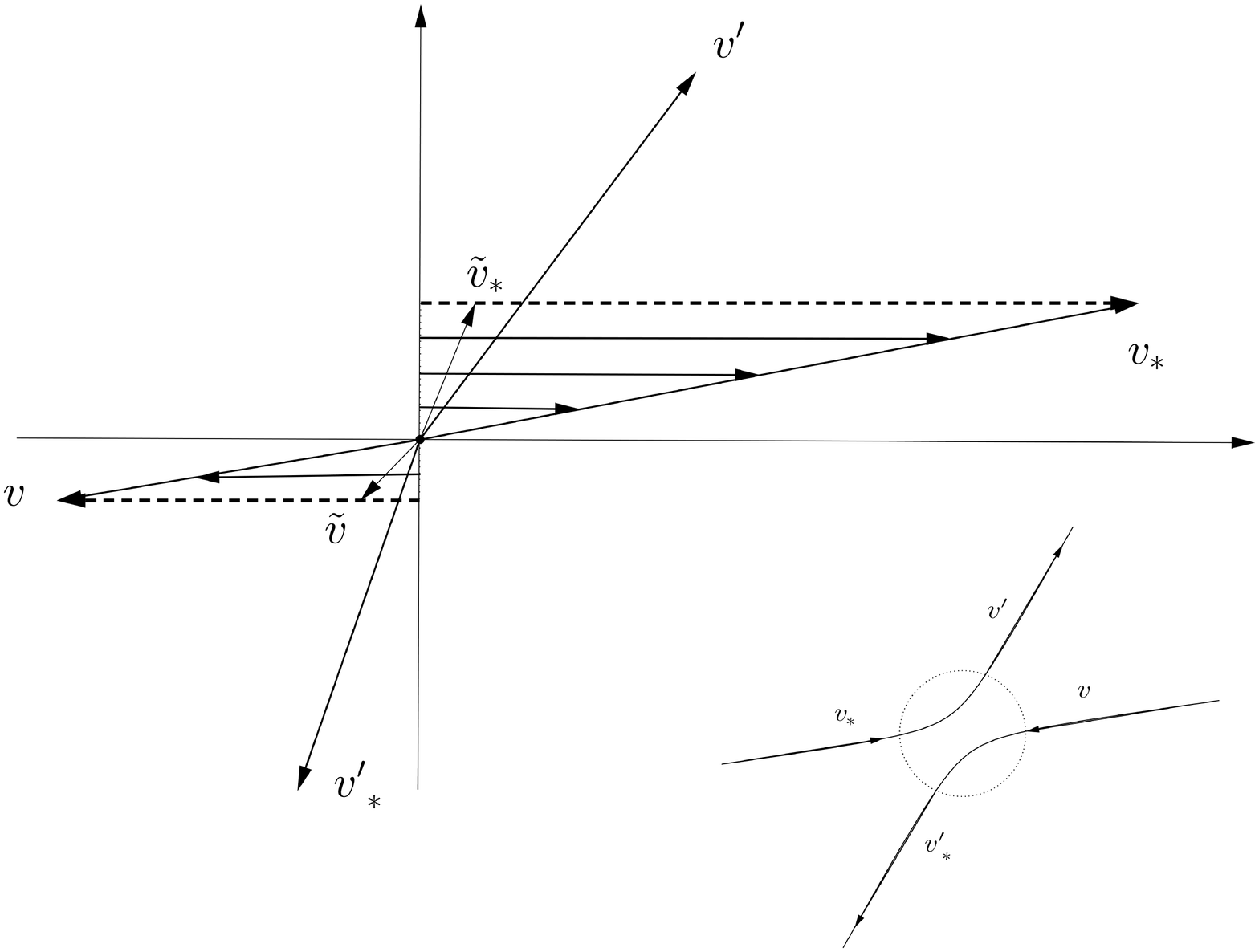}\caption{Increase of average
velocity due to the combined effect of shear and collisions. The velocities
$\tilde{v}, \tilde{v}_{*}$ are transformed into $v, {v}_{*}$ by the shear.
Then the collisions transform these velocities into $v^{\prime}, {v^{\prime}%
}_{*}$. }%
\label{fig:ShearColl}%
\end{figure}

A feature that characterizes several of the homoenergetic flows dominated by hyperbolic
terms is the fact that the collisions term, in spite of the fact that it is
formally very small as $t\rightarrow\infty$, yields huge effects in the
particle distributions. This feature can be understood in an intuitive manner
in terms of the trajectories of the particles which are described by the
Boltzmann equation in homoenergetic flows.
During most of the time the dynamics of the particles is described by the
hyperbolic flow, typically a shear flow, a dilatation flow or a combination of
them. Rarely, the particle experiences a collision with other particles and
this modifies drastically the direction of the motion of the particle. Then
the particle velocity evolves again according to the hyperbolic flows and this
results in an additional increase of the size of the velocities.
Therefore the iteration of this process yields a huge increase of the average
velocities and therefore of the ``temperature" of the system. Moreover, the
effect of the collisions in the long time asymptotics of the particle distribution is huge in spite of
the fact that they take place very rarely. See Figure \ref{fig:ShearColl}. This phenomenon will be studied in Subsection \ref{sec:domhyper}  relying on the analysis of a collisional model simpler than the Boltzmann equation.

\bigskip

 As discussed above the information that we have obtained so far for these
flows with dominant hyperbolic terms is less detailed and more fragmentary than the one that we have obtained in the collision-dominated case. We will
describe below a few examples of these flows and we will describe the dynamics
of a simplified model which contains the combined effect
explained above (simple shear during large times combined with rare
collisions). Hopefully some of the ideas described in such simplified model
might be useful to understand homoenergetic flows for the Boltzmann equation
dominated by hyperbolic terms.

\subsection{Homogeneous dilatation: Frozen collisions}

\label{ss:3ddilslow}

\bigskip

We recall that the homoenergetic flows (\ref{B1_0}) in the case of homogeneous
dilatation (i.e. $L\left(  t\right)  $ given by (\ref{T1E1})) have been
studied in \cite{Nikol1}, \cite{Nikol2}.  We observe that in these flows the
average dispersion of the velocity of the particles decreases as $\frac{1}{t}$ as $t\rightarrow
\infty,$ something that it is just due to the dilatation process of the gas
with the collisions not having any meaningful effect on this fact.  The long
time asymptotics of the distribution then depends on the behavior of the
average time between collisions and this depends on the homogeneity of the
collision kernel $B$ which we denote as $\gamma.$ As it has been seen in \cite{JNV2} if $\gamma\leq-2$ the effect of the collisions
is relevant in the long time asymptotics and the velocity distribution
converges to a Maxwellian distribution. If $\gamma>-2$,  we will obtain the behavior that we will denote as frozen collisions as $t\to \infty$. Indeed, the effect of the
isotropic dilatation is to reduce the average dispersion of the velocity of the molecules.
Therefore, the effect of the collisions becomes negligible for large times and
the particle distribution $g\left(  t,w\right)  $ converges asymptotically to
a distribution $g_{\infty}\left(  w\right)  $ which depends on the initial
particle distribution $g_{0}\left(  w\right)  .$

To be more precise, we consider the following equation which has been derived in Subsection 4.2.4 in \cite{JNV2} 
\begin{equation}
\label{T6E7_1}\partial_{\tau}G-\partial_{\xi}\cdot\left(  \bar{\alpha}\left(
\tau\right)  \xi G\right)  =e^{-\left(  2+\gamma\right)  \tau}\mathbb{C}%
G\left(  \xi\right)
\end{equation}
where $\left\vert \bar{\alpha}\left(  \tau\right)  \right\vert \leq Ce^{-\tau
}.$ Looking at the new time scale
\[
ds=e^{-\left(  2+\gamma\right)  \tau}d\tau \
\]
when $\gamma>-2$, we then have $\left(  2+\gamma\right)  >0$ and
\[
s=\frac{1}{\left(  2+\gamma\right)  }\left(  1-e^{-\left(  2+\gamma\right)
\tau}\right)  .
\]
Therefore (\ref{T6E7_1}) becomes
\begin{equation}
\partial_{s}G-\partial_{\xi}\cdot\left(  \kappa\left(  s\right)  \xi G\right)
=\mathbb{C}G\left(  \xi\right)  \label{T6E8}%
\end{equation}
where $s$ takes value in the interval $\left[ 0,\frac{1}{(  2+\gamma)}\right)$ when $\tau$ takes value in $[0,\infty)$.
  Moreover,  
$\kappa\left(  s\right)  $ is bounded in the interval $0\leq s\leq \frac{1}{(  2+\gamma)}.$
 Therefore, in order to compute the asymptotic behaviour of the solutions of \eqref{T6E7_1} as $\tau\rightarrow\infty$, we just need to compute the asymptotic behaviour of the solutions $G(s,w)$ of (\ref{T6E8}) as $s\to \frac
{1}{\left(  2+\gamma\right)  } .$   Therefore $G\left(  s,w\right)  $
does not converge to a Gaussian, but to a limit measure $G_{\infty}\left(
\xi\right)  $ which depends on the initial value $g_{0}\left(  w\right)  .$

\subsection{Cylindrical dilatation: Frozen collisions}

We now consider \eqref{D1_0} choosing $L(t)$ as in \eqref{T1E2} with $K=0$. We obtain the following equation 
\begin{equation}
\partial_{t}g-\frac{1}{t}\left(  w_{1}\partial_{w_{1}}+w_{2}\partial_{w_{2}%
}\right)  g=\int_{\mathbb{R}^{3}}dw_{\ast}\int_{S^{2}}d\omega\,B\left(
\omega,\left\vert w-w_{\ast}\right\vert \right)  \left[  g^{\prime}g_{\ast
}^{\prime}-g_{\ast}g\right]  \label{eq:Boltz2dilHilb1}%
\end{equation}
when the kernel has homogeneity
$\gamma>-2.$ We remark that the case of Maxwellian molecules (i.e.~$\gamma=0$) has been considered by Galkin in \cite{Galkin3} where formulas for
the second order moments of the velocity distribution have been computed using
hypergeometric functions.

In order to rewrite the equation above in divergence form we use the change of variables:  $g(t,w)=\frac
{1}{t^{2}}G(t,w)$, $\log(t)=\tau$. Then, \eqref{eq:Boltz2dilHilb1} becomes 
\begin{align}
\label{eq:2ddileq}
\partial_{\tau}G-\partial_{w}\cdot\left(  Dw G\right)  =e^{-\tau} \mathbb{C}G,\;\; \text{with} \;\; 
D= \left(
\begin{array}
[c]{ccc}%
1 & 0 & 0\\
0 & 1 & 0\\
0 & 0 & 0
\end{array}
\right).  
\end{align}

We study under which conditions we can obtain solutions of \eqref{eq:2ddileq} 
whose behavior is determined for $\tau\rightarrow\infty$ by
the terms on the left-hand side.  Suppose that we replace the collision term $e^{-\tau}\mathbb{C}G$ on the right-hand side of
\eqref{eq:2ddileq} by $0$. Then, the solutions by the method of characteristics of the corresponding equation would be given by
\begin{equation}
G\left(  \tau,w\right)  =e^{2\tau}G_{0}\left(  e^{\tau}w_{1},e^{\tau}%
w_{2},w_{3}\right)  \label{D3E1}%
\end{equation}
where $G_{0}$ is the initial distribution which we assume to be sufficiently  smooth and exponentially decaying as $|w|\to \infty$.

We now check that the collision terms obtained with a function $G$ of the form \eqref{D3E1} will not modify the form of the solutions obtained by the method of characteristics as $\tau\to \infty$.    
The \emph{rate of collisions} is given by%
\[
e^{-\tau}\int_{\mathbb{R}^{3}}dw_{\ast}\int_{S^{2}}d\omega B\left(n \cdot  \omega
,\left\vert w-w_{\ast}\right\vert \right)  G\left(  \tau,w_{\ast}\right)
\]
which can be estimated as
\begin{equation}
Ce^{-\tau}\int_{\mathbb{R}^{3}}dw_{\ast}\left\vert w-w_{\ast}\right\vert
^{\gamma}G\left(  \tau,w_{\ast}\right)  . \label{D3E2}%
\end{equation}

We estimate the supremum in $w$ of (\ref{D3E2}). If we assume that $G$ has the
form (\ref{D3E1}) we can estimate this supremum, for $\vert w\vert \leq C$, by the value at $w=0.$ We then
need to estimate
\begin{equation}
Ce^{-\tau}\int_{\mathbb{R}^{3}}dw_{\ast}\left\vert w_{\ast}\right\vert
^{\gamma}G\left(  \tau, w_{\ast}\right)  . \label{D3E3}%
\end{equation}

Due to the assumption we did on the initial distribution $G_0$ it follows that $G$ as well as the moment
$\int_{\mathbb{R}^{3}}dw_{\ast}\left\vert w_{\ast}\right\vert ^{\left\vert
\gamma\right\vert }G\left(  w_{\ast},\tau\right)  <\infty$ for each $\tau
\geq0$. We consider separately the cases $\gamma\geq0$ and $\gamma<0.$ If
$\gamma\geq0$ and $G$ has the form (\ref{D3E1}) we obtain that
(\ref{D3E3}) is bounded by $Ce^{-\tau},$ since the mass of $G$ is concentrated
in the region where $\left\vert w_{\ast}\right\vert $ is bounded. Therefore,
the rate of collisions decreases exponentially as $\tau\rightarrow\infty$ and
we can expect the asymptotics (\ref{D3E1}) for $G$ as $\tau\rightarrow\infty.$
Suppose now that $-2<\gamma<0$ and that $G$ has the form (\ref{D3E1}). We then
obtain, using the change of variables $\xi_1=e^{\tau}w_{1},\, \xi_2=e^{\tau}w_{2}$,  the following estimate for (\ref{D3E3})%
\[
Ce^{-\tau}\int_{\mathbb{R}^{3}}d\xi_{1}d\xi_{2}dw_{3}\left[  e^{-2\tau}\left(
\left(  \xi_{1}\right)  ^{2}+\left(  \xi_{2}\right)  ^{2}\right)  +\left(
w_{3}\right)  ^{2}\right]  ^{\frac{\gamma}{2}}G_{0}\left(  \xi_{1},\xi
_{2},w_{3}\right).
\]
We assume that $G_{0}$ satisfies \eqref{D3E1} 
and contains most of its mass in the
region $\left\vert \xi_{1}\right\vert +\left\vert \xi_{2}\right\vert
+\left\vert w_{3}\right\vert \leq C$. Performing then the change of variables $w_3 =e^{-\tau} \eta \vert \xi \vert $ with $\xi=\left(  \xi_{1},\xi_{2}\right)  \in\mathbb{R}^{2}$, we obtain an estimate with the form%
\[
Ce^{-(2-\vert\gamma\vert)\tau}\int_{B_{A}\left(  0\right)  }d\xi \vert \xi \vert ^{1-\vert\gamma\vert}\int_{-\frac{Ae^{\tau}}{\left\vert
\xi\right\vert }}^{\frac{Ae^{\tau}}{\left\vert \xi\right\vert }}\frac{d\eta
}{\left[  1+\eta^{2}\right]  ^{\frac{\left\vert \gamma\right\vert }{2}}}%
\] 
where $A>0.$ 

We
now have two possibilities. If 
$1<\left\vert \gamma\right\vert <2$, $\gamma<0$, the integral above is bounded by $Ce^{-(2-\vert\gamma\vert)\tau}.$  If $\left\vert \gamma\right\vert <1$, $\gamma<0$, we
obtain the estimate
\[
Ce^{-\tau}\int_{B_{A}\left(  0\right)  } 
d\xi\leq
Ce^{-\tau}.
\]
If $\left\vert \gamma\right\vert =1$ we would obtain the estimate $C\tau e^{-\tau}$ with similar arguments. 
Summarizing, if $\gamma>-2$ the collision rate decreases exponentially and
we can expect to have an asymptotics for $G$ given by
\begin{equation}
\label{eq:Ginfinity}G\left(  \tau, w\right)  =e^{2\tau}G_{\infty}\left(
e^{\tau}w_{1},e^{\tau}w_{2},w_{3}\right)
\end{equation}
where $G_{\infty}$ would depend on $G_{0}.$

\begin{remark}
It is interesting to remark that in \cite{JNV2} we have obtained that in the case of cylindrical dilatation with kernels $B$ with
homogeneity $\gamma<-\frac{3}{2}$ there are homoenergetic solutions for the
Boltzmann equation described by means of Hilbert expansions and behaving like
a Maxwellian distribution with decreasing temperature. On the other hand we
have obtained in this subsection that a possible asymptotics of the
homoenergetic solutions for the Boltzmann equation is given by
(\ref{eq:Ginfinity}) if $\gamma>-2.$ The remarkable fact is that there is a non
empty interval of homogeneities $\gamma\in\left(  -2,-\frac{3}{2}\right)  $
for which both asymptotics are possible. This suggests that the homoenergetic
solutions of the Boltzmann equation can have different behaviors depending on
the choice of initial data $G_{0}$ if $\gamma\in\left(  -2,-\frac{3}%
{2}\right).$  In one of the asymptotics the effect of the collisions would be
the dominant effect, while in the other one, the collisions would have a
negligible effect for large times. This is a remarkable phenomenon which
deserves a more detailed analysis, nonetheless we will not continue with the
study of this case in this paper. \bigskip
\end{remark}

\subsection{Simple shear. \  Frozen
collisions for $\gamma<-1$.  \label{SimpShearFrColl}}

We consider homoenergetic flows \eqref{D1_0} with $L\left(  t\right)  $ as in
(\ref{T1E5}). Then $g$ satisfies:%
\begin{equation}
\partial_{t}g-Kw_{2}\partial_{w_{1}}g=\mathbb{C}g\left(  w\right)  .
\label{A1E1}%
\end{equation}

We will show that if the homogeneity of the collision kernel $B$ is smaller
than $-1$ there exist solutions of (\ref{A1E1}) for which the contribution of
the collision term $\mathbb{C}g\left(  w\right)  $ is negligible as
$t\rightarrow\infty.$ For such solutions, a given particle would
not collide with any other for large times, and we might expect to have $w_{2}$
approximately constant and $w_{1}$ increasing linearly in $t.$ This
suggests to look for solutions with the form
\begin{equation}
g\left(  t,w\right)  =\frac{1}{t}G\left(  \tau,\xi\right)  \ \ ,\ \ \tau
=\log\left(  t\right)  ,\ \ \xi_{1}=\frac{w_{1}}{t},\ \xi_{j}=w_{j}\text{ if
}j=2,3 .\label{A3E1}%
\end{equation}

Then, using the homogeneity of the kernel, we obtain that $G$ satisfies%
\begin{equation}
\partial_{\tau}G-\partial_{\xi}\cdot\left(  \left[  \left(  \xi_{1}+K\xi
_{2}\right)  e_{1}\right]  G\right)  =e^{\left(  1+\gamma\right)  \tau
}\mathbb{C}G\left(  \xi\right)  . \label{A3E2}%
\end{equation}

Notice that the collision kernel has been also rescaled, due to the lack of isotropy of the change of variables \eqref{A3E1}. 

The long time asymptotics of the hyperbolic equation obtained putting
the right-hand side equal to zero in \eqref{A3E2} is much dependent on the regularity properties of
the initial data $G_{0}\left(  \xi\right)  .$ We will assume for the sake of
simplicity that $G_{0}\in C^{2}.$ The solution of the corresponding hyperbolic
equation follows using the method of characteristics%
\begin{align}
G\left(  \tau,\xi\right)   &  =e^{\tau}G_{0}\left(  \xi_{1}e^{\tau}+K\xi
_{2}\left(  e^{\tau}-1\right)  ,\xi_{2},\xi_{3}\right), \label{A3E3}\\
g\left(  t,w\right)  &  =t\, G_{0}\left(  w_{1}+Kw_{2}\left(  t-1\right)  ,w_{2},w_{3}\right)
.\nonumber
\end{align}

Suppose that $G_{0}$ is compactly supported or decreases sufficiently fast as
$\vert w \vert \rightarrow\infty.$ Then, integrating with respect to $\xi$ against a smooth test function we obtain
\begin{equation}
G\left(  \tau,\xi\right)  \rightharpoonup\left[  \int_{-\infty}^{\infty}%
G_{0}\left(  \eta,\xi_{2},\xi_{3}\right)  d\eta\right]  \delta\left(  \xi
_{1}+K\xi_{2}\right)  \ \ \text{as\ \ }\tau\rightarrow\infty. \label{A3E4}%
\end{equation}

If $\gamma<-1$ it would follow that the contribution of the collision term
$e^{\left(  1+\gamma\right)  \tau}\mathbb{C}G\left(  \xi\right)  $ decreases
exponentially as $\tau\rightarrow\infty$ and it yields a negligible
contribution as $\tau\rightarrow\infty.$ In this case the effect of the
collisions becomes frozen for large times. The rigorous proof of the smallness
of this term would require some careful analysis of the collision term which
will not be made in detail in this paper. The intuitive idea behind the
convergence (\ref{A3E4}) is that collisions, for long times, take place so rarely that become negligible.

\subsection{Combined shear in orthogonal directions \label{CombShear}}

\subsubsection{Derivation of a system of ODEs for the second order moments}

We now consider the homoenergetic flows (\ref{D1_0}) with $L\left(  t\right)
$ as in (\ref{T1E4}). Then $g$ solves

\begin{equation}
\partial_{t}g-\left[  K_{3}w_{2}+\left(  K_{2}-tK_{1}K_{3}\right)
w_{3}\right]  \partial_{w_{1}}g-K_{1}w_{3}\partial_{w_{2}}g=\mathbb{C}g\left(
w\right) . \label{T6E9}
\end{equation}

It readily follows that $\partial_{t}\left(  \int_{\mathbb{R}^{3}}g\left(
t,dw\right)  \right)  =0.$ On the other hand, as we have seen in \cite{JNV2}, for homogeneity $\gamma>0$ the asymptotic behaviour of solutions of (\ref{T6E9}) is given by Maxwellian distributions with time dependent temperature, obtained by means of suitable Hilbert expansions. Such
expansions do not exist for $\gamma<0.$ Therefore we can expect the critical
value of the homogeneity at $\gamma=0.$ We can obtain some insight about the
asymptotics of $g$ from the asymptotics of the second moments $M_{j,k}=\int_{\mathbb{R}^{3}}w_{j}w_{k}g\left(  t,dw\right).$ 
The asymptotic formulas we obtain for the moments will rule out the possibility of self-similar behaviour.

We consider the evolution equation for the second moments $M_{j,k}=\int_{\mathbb{R}^{3}}w_{j}w_{k}g\left(  t,dw\right).$   
From \eqref{T6E9} at $\gamma=0$ a straightforward computation yields
\begin{align}
  \frac{dM_{j,k}}{dt}&=K_{3}\delta_{j,1}M_{k,2}+\left[  \left(  K_{2}%
-tK_{1}K_{3}\right)  \delta_{j,1}+K_{1}\delta_{j,2}\right]  M_{k,3}%
+\nonumber\\
&\quad  +K_{3}\delta_{k,1}M_{j,2}+\left[  \left(  K_{2}-tK_{1}K_{3}\right)
\delta_{k,1}+K_{1}\delta_{k,2}\right]  M_{j,3}\nonumber\\
& \quad -  2b\left(  M_{j,k}-m\delta_{j,k}\right)  \ \label{T7E2}%
\end{align}
for $j,k=1,2,3,\ \ \ M_{j,k}=M_{k,j},\ m=\frac{1}{3}\left(  M_{1,1}%
+M_{2,2}+M_{3,3}\right)  $  and 
\begin{equation}
b=3\pi\int_{-1}^{1}B\left(  x\right)  x^{2}\left(  1-x^{2}\right)  dx>0.
\label{eq:b}%
\end{equation}

\medskip

The system of equations (\ref{T7E2}) is linear. Due to the assumption
$K_{1}K_{3}\neq0$ there is a linear increase
of the terms on the right-hand side of this system. It is then natural to
derive asymptotic formulas for its solutions using a WKB method (see for instance \cite{BeOr}). Indeed, these
methods are applicable for linear problems for which the terms multiplying the higher order derivatives are asymptotically smaller than the terms multiplying lower order derivatives. We will derive asymptotic formulas for six linearly independent
solutions of (\ref{T7E2}). We make the ansatz $M_{j,k}\sim\gamma_{j,k}e^{S}$
where $\gamma_{j,k}$ are functions behaving like power laws (perhaps
containing logarithmic corrections) and $S$ behaves like a polynomial. We
first need to compute the leading order of this asymptotic expansion. To this
end we rewrite (\ref{T7E2}) neglecting the contribution of the term $K_{2}$ in
$\left(  K_{2}-tK_{1}K_{3}\right)$ since it is lower order. With this approximation we have
\begin{align}
\frac{dM_{1,1}}{dt}  &  =-2K_{3}M_{1,2}+2tK_{1}K_{3}M_{1,3}-2b\left(
M_{1,1}-m\right) \nonumber\\
\frac{dM_{1,2}}{dt}  &  =-K_{3}M_{2,2}+tK_{1}K_{3}M_{2,3}-K_{1}M_{1,3}%
-2bM_{1,2}\nonumber\\
\frac{dM_{1,3}}{dt}  &  =-K_{3}M_{2,3}+tK_{1}K_{3}M_{3,3}-2bM_{1,3}\nonumber\\
\frac{dM_{2,2}}{dt}  &  =-2K_{1}M_{2,3}-2b\left(  M_{2,2}-m\right) \nonumber\\
\frac{dM_{2,3}}{dt}  &  =-K_{1}M_{3,3}-2bM_{2,3}\nonumber\\
\frac{dM_{3,3}}{dt}  &  =-2b\left(  M_{3,3}-m\right) \label{T7E3}\\
m  &  =\frac{1}{3}\left(  M_{1,1}+M_{2,2}+M_{3,3}\right)  .\nonumber
\end{align}

\subsubsection{Computations of the asymptotics for the second order moments}

\begin{figure}[th]
\centering
\includegraphics [scale=0.2]{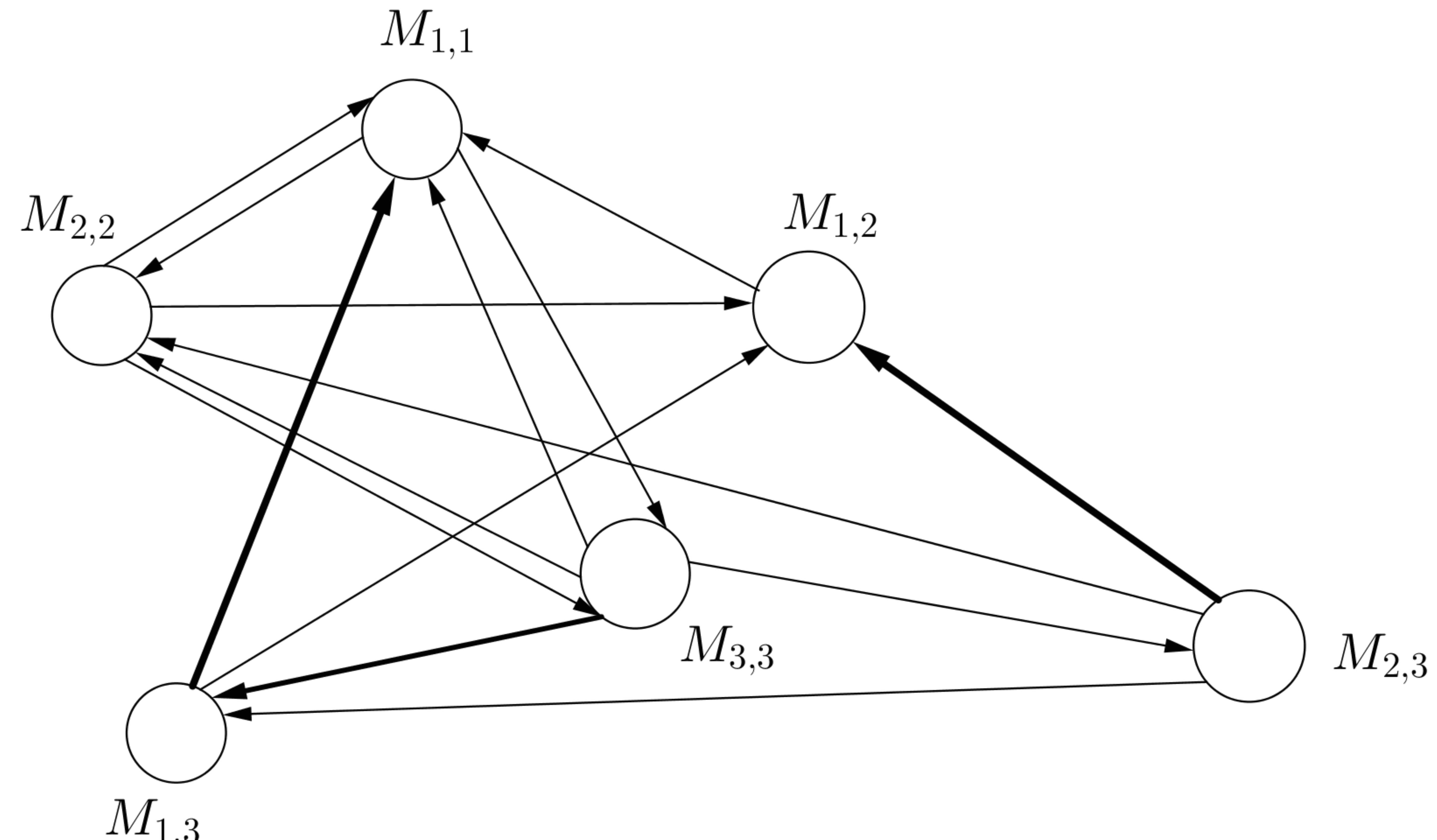}
\caption{The graphic procedure for the WKB method.}\label{fig:WKB} 
\end{figure}

We can determine the exponential behavior of the coefficients using a graphic
procedure. To this end we represent each of the variables $M_{j,k}$ as the
nodes of a graph. (See Figure \ref{fig:WKB}). 
We then include in the graph a
directed edge connecting each of the nodes appearing on the right-hand side of
(\ref{T7E3}) (including those on $m$) with the variable appearing under the
derivative on the right-hand side. We will assume that the directed edges are
of two different types, namely thick and thin. More precisely, we add an edge
of type thin if the term appearing on the right-hand side is multiplied by a
constant, and we will assume that the edge is thick if the corresponding
variable on the right-hand side of (\ref{T7E3}) is proportional to $t.$
Suppose that $M_{j,k}$ is a variable at the origin of one directed edge on the
graph and $M_{r,s}$ is a variable at the end of one of such directed edges.
Then, the structure of the equations (\ref{T7E3}) implies the following rule
to determine the structure of the algebraic factors $\gamma_{j,k}$%
\begin{align}
c_{j,k}\gamma_{j,k}  &  =\left(  \partial_{t}S\right)  \gamma_{r,s}%
\ \ \text{if the directed edge is thin},\label{T7E4}\\
c_{j,k}t\gamma_{j,k}  &  =\left(  \partial_{t}S\right)  \gamma_{r,s}%
\ \ \text{if the directed edge is thick},\nonumber
\end{align}
where $c_{j,k}$ is the multiplicative constant in front of the function
$M_{j,k}$ in (\ref{T7E3}).

We then look for circuits in the graph associated to the set of equations
(\ref{T7E3}) having the shortest length and the largest number of thick lines.
More precisely, for any circuit connecting a node $M_{j,k}$ with itself, the
set of rules (\ref{T7E4}) implies that%
\begin{equation}
C_{0}\gamma_{j,k}t^{T}=\left(  \partial_{t}S\right)  ^{L}\gamma_{j,k}
\label{T7E5}%
\end{equation}
where $L$ is the total number of edges of the circuit and $T$ is the number of
thick lines on it. The constant $C_{0}$ is just the product of the
coefficients $c_{j,k}$ in the cycle.

We select the circuits for which the number $\frac{T}{L}$ is the largest.
Notice that by construction $\frac{T}{L}<1.$ Then, the consistency of
(\ref{T7E5}) yields
\begin{equation}
S\sim\frac{\left\vert C_{0}\right\vert ^{\frac{1}{L}}\omega}{1+\frac{T}{L}%
}t^{1+\frac{T}{L}}\text{ as }t\rightarrow\infty\label{T7E6}%
\end{equation}
where $\omega$ is one of the $L$ complex roots of the equation
\[
\omega^{L}=\operatorname*{sgn}\left(  C_{0}\right)  .
\]

The asymptotics (\ref{T7E6}) yields the leading asymptotic behavior of $L$
independent solutions of (\ref{T7E3}). We can derive asymptotic formulas for additional solutions removing the nodes which are at the basis of the thick lines contained in the cycles yielding contributions to the asymptotics (\ref{T7E6}), if needed. In such a case, in order to obtain the asymptotics of the additional solutions, we obtain formulas relating some of the asymptotic variables $M_{j,k}$ removing the derivatives in the equations associated to the thick lines removed and finding then asymptotic relations between the corresponding right-hand sides of the resulting equations.

We apply the method explained above to the equations (\ref{T7E3}). The cycle
yielding the smallest value of $\frac{T}{L}$ is
\begin{equation}
M_{1,1}\rightarrow M_{3,3}\rightarrow M_{1,3}\rightarrow M_{1,1} .
\label{U1E1}%
\end{equation}

Then, we obtain to the leading order
\[
\left(  S_{t}\right)  ^{3}\sim\frac{4b}{3}\left(  tK_{1}K_{3}\right)
^{2}\text{ as }t\rightarrow\infty
\]
whence
\begin{equation}
S\sim\frac{3}{5}\left(  \frac{4b}{3}\right)  ^{\frac{1}{3}}\left(  K_{1}%
K_{3}\right)  ^{\frac{2}{3}}\omega t^{\frac{5}{3}}\ \ \text{as }%
t\rightarrow\infty\label{T7E8}%
\end{equation}
where $\omega^{3}=1.$ We are interested in the value of $\omega$ yielding the
fastest growth of the solutions, i.e. $\omega=1.$

We compute the remaining long time asymptotics of (\ref{T7E3}) as a
consistency test. We will impose that $\frac{dM_{3,3}}{dt}=\frac{dM_{1,3}}%
{dt}=\frac{dM_{1,1}}{dt}=0$ (or, more precisely, that the contributions of these derivatives yield subdominant
terms). We then need to solve the equations%
\begin{align}
\frac{dM_{1,2}}{dt}  &  =-K_{3}M_{2,2}+tK_{1}K_{3}M_{2,3}-K_{1}M_{1,3}%
-2bM_{1,2}\label{T7E7}\\
\frac{dM_{2,2}}{dt}  &  =-2K_{1}M_{2,3}-2b\left(  M_{2,2}-m\right) \nonumber\\
\frac{dM_{2,3}}{dt}  &  =-K_{1}M_{3,3}-2bM_{2,3}\ \nonumber\\
m  &  =\frac{1}{3}\left(  M_{1,1}+M_{2,2}+M_{3,3}\right) \nonumber
\end{align}
with the constraints
\begin{align*}
0  &  =-2K_{3}M_{1,2}+2tK_{1}K_{3}M_{1,3}-2b\left(  M_{1,1}-m\right) \\
0  &  =-K_{3}M_{2,3}+tK_{1}K_{3}M_{3,3}-2bM_{1,3}\\
0  &  =-2b\left(  M_{3,3}-m\right)  .
\end{align*}

Using the last equation we can write 
$$\left(  M_{1,1}-m\right)  =\left(
M_{1,1}-M_{3,3}\right)  +\left(  M_{3,3}-m\right)  =\left(  M_{1,1}%
-M_{3,3}\right)  ,$$ whence
\begin{align}
0  &  =-2K_{3}M_{1,2}+2tK_{1}K_{3}M_{1,3}-2b\left(  M_{1,1}-M_{3,3}\right)
\nonumber\\
0  &  =-K_{3}M_{2,3}+tK_{1}K_{3}M_{3,3}-2bM_{1,3}\nonumber\\
0  &  =2M_{3,3}-M_{1,1}-M_{2,2} . \label{T7E7a}%
\end{align}

Neglecting subdominant terms and using the first equation we obtain
\begin{equation}
M_{1,3}=\frac{K_{3}M_{1,2}+bM_{1,1}}{K_{1}K_{3}}\frac{1}{t} . \label{T7E7b}%
\end{equation}
Plugging this into the second equation of (\ref{T7E7a}) we obtain
\begin{equation}
M_{3,3}=\frac{M_{2,3}}{K_{1}}\frac{1}{t}+\frac{2b\left(  K_{3}M_{1,2}%
+bM_{1,1}\right)  }{\left(  K_{1}K_{3}\right)  ^{2}}\frac{1}{t^{2}}.
\label{T7E7c}%
\end{equation}

Notice that we have neglected the term $-\frac{bM_{3,3}}{K_1K_3t}$ in \eqref{T7E7b}. This would result in a term of order $O\left(\frac{M_{3,3}}{t^3}\right)$ in \eqref{T7E7c} which is lower oder $M_{3,3}$ compared to the term on the left of \eqref{T7E7c}. Therefore, we would neglect it.  
Inserting now this formula into the last equation of (\ref{T7E7a}) and
neglecting small terms we obtain
\[
\frac{2M_{2,3}}{K_{1}}\frac{1}{t}+\frac{4b\left(  K_{3}M_{1,2}+bM_{1,1}%
\right)  }{\left(  K_{1}K_{3}\right)  ^{2}}\frac{1}{t^{2}}-M_{1,1}-M_{2,2}=0
\]%
\begin{equation}
M_{1,1}=\frac{2M_{2,3}}{K_{1}}\frac{1}{t}+\frac{4bK_{3}M_{1,2}}{\left(
K_{1}K_{3}\right)  ^{2}}\frac{1}{t^{2}}-M_{2,2} . \label{T7E8a}%
\end{equation}

Using this into (\ref{T7E7b}), (\ref{T7E7c}) and neglecting small terms we
get

\begin{equation}
M_{1,3}=\left(  \frac{M_{1,2}}{K_{1}}\frac{1}{t}+\frac{2bM_{2,3}}{\left(
K_{1}\right)  ^{2}K_{3}}\frac{1}{t^{2}}-\frac{bM_{2,2}}{K_{1}K_{3}}\frac{1}%
{t}\right)  \ \label{T7E8b}%
\end{equation}

\begin{equation}
M_{3,3}=\frac{M_{2,3}}{K_{1}}\frac{1}{t}+\frac{2bK_{3}M_{1,2}}{\left(
K_{1}K_{3}\right)  ^{2}}\frac{1}{t^{2}}+\frac{2b^{2}M_{2,2}}{\left(
K_{1}K_{3}\right)  ^{2}}\frac{1}{t^{2}}. \label{T7E8c}%
\end{equation}

\bigskip

We can now use (\ref{T7E8a})-(\ref{T7E8c}) to eliminate $M_{1,1}%
,M_{1,3},M_{3,3}$ from (\ref{T7E7}). Eliminating subdominant terms we
obtain
\begin{align}
\frac{dM_{1,2}}{dt}  &  =-K_{3}M_{2,2}+tK_{1}K_{3}M_{2,3}-2bM_{1,2}\\
\frac{dM_{2,2}}{dt}  &  =-2K_{1}M_{2,3}-2bM_{2,2}+\frac{4b^{2}K_{3}M_{1,2}%
}{3\left(  K_{1}K_{3}\right)  ^{2}}\frac{1}{t^{2}}\nonumber\\
\frac{dM_{2,3}}{dt}  &  =\frac{2bK_{3}M_{1,2}}{\left(  K_{1}K_{3}\right)
^{2}}\frac{1}{t^{2}}-K_{1}\frac{2b^{2}M_{2,2}}{\left(  K_{1}K_{3}\right)
^{2}}\frac{1}{t^{2}}-2bM_{2,3}\ .\nonumber
\end{align}

These equations have three independent solutions which decrease exponentially. More precisely, the asymptotics of these solutions can be computed with the same type of arguments used above and the conclusion is the existence of three  independent solutions having the following asymptotics. Two of them behave like:
\begin{align}
M_{j,k}=\exp\left(-2bt\pm \sqrt{\frac{2b}{K_1}t}\,(1+\ep_{j,k}(t))\right)\quad \text{for}\quad (j,k)\in\{(1,2),(2,2),(2,3)\}  \quad\text{as}\quad t\to\infty
\end{align}
where $\ep_{j,k}(t)\to 0$ as $t\to\infty$,  and the third one behaves like:
\begin{align}
M_{j,k}=e^{-2bt} \gamma_{j,k}\quad \text{for}\quad (j,k)\in\{(2,3)\} \quad (j,k)\in\{(1,2),(2,2),(2,3)\} \quad\text{as}\quad t\to\infty
\end{align}
where $\gamma_{j,k}$ are such that $\gamma_{2,2}=\frac{1}{t^2}$, $\gamma_{2,3}\sim \frac{1}{K_1\,t}\gamma_{2,2}$, $\gamma_{1,2}\sim \frac{K_1\,b}{K_3}\gamma_{2,2}$ as $t\to\infty$.
This means that we have three additional independent solutions to the ones having the
asymptotics (\ref{T7E8}), but we will not give more details about them in what follows since we will not use them later.

\medskip

The relevant asymptotics, which yields the behavior of the functions $M_{j,k}$
for generic initial data is the one yielding $S\sim\frac{3}{5}\left(\frac{  4b}{3}\right)
^{\frac{1}{3}}\left(  K_{1}K_{3}\right)  ^{\frac{2}{3}}t^{\frac{5}{3}}$ as
$t\rightarrow\infty.$ We can compute additional terms in the asymptotics of
the solutions $M_{j,k}.$

To this end we introduce the change of variables
\begin{equation}
M_{j,k}=e^{S_{0}+H_{j,k}}\ \ ,\ \ S_{0}=\frac{3}{5}\left(  \frac{4b}%
{3}\right)  ^{\frac{1}{3}}\left(  K_{1}K_{3}\right)  ^{\frac{2}{3}}t^{\frac
{5}{3}} \label{T7E9}%
\end{equation}
where the behavior of the functions $H_{j,k}$ must be computed. By assumption
$\vert H_{j,k}\vert \ll t^{\frac{5}{3}}$ as $t\rightarrow\infty.$ We will assume also that
terms in the differential equations which are not associated to the nodes
appearing in the cycle (\ref{U1E1}) yield negligible contributions. We will
check then ``a posteriori" \ that these assumptions are satisfied in the
derived asymptotic formulas for $H_{j,k}.$ We remark that the leading order
asymptotics derived above imply
\begin{equation}
e^{H_{1,3}-H_{1,1}}\sim\frac{\partial_{t}S_{0}}{2tK_{1}K_{3}}\ ,\ \ e^{H_{3,3}%
-H_{1,3}}\sim\frac{\partial_{t}S_{0}}{tK_{1}K_{3}}\ ,\ \ e^{H_{1,1}-H_{3,3}%
}\sim\frac{3\partial_{t}S_{0}}{2b} . \label{U1E3}%
\end{equation}

On the other hand, plugging (\ref{T7E9}) into (\ref{T7E3}) we obtain
\begin{align}
\partial_{t}S_{0}+\partial_{t}H_{1,1}  &  =2tK_{1}K_{3}e^{H_{1,3}-H_{1,1}%
}-2K_{3}e^{H_{1,2}-H_{1,1}}-K_{2}e^{H_{1,3}-H_{1,1}}-\frac{2b}{3}\left(
2-e^{H_{2,2}-H_{1,1}}-e^{H_{3,3}-H_{1,1}}\right) \nonumber\\
\partial_{t}S_{0}+\partial_{t}H_{1,2}  &  =-K_{3}e^{H_{2,2}-H_{1,2}}+\left(
tK_{1}K_{3}-K_{2}\right)  e^{H_{2,3}-H_{1,2}}-K_{1}e^{H_{1,3}-H_{1,2}%
}-2b\nonumber\\
\partial_{t}S_{0}+\partial_{t}H_{1,3}  &  =tK_{1}K_{3}e^{H_{3,3}-H_{1,3}%
}-K_{3}e^{H_{2,3}-H_{1,3}}-K_{2}e^{H_{3,3}-H_{1,3}}-2b\nonumber\\
\partial_{t}S_{0}+\partial_{t}H_{2,2}  &  =-2K_{1}e^{H_{2,3}-H_{2,2}}%
-\frac{2b}{3}\left(  2-e^{H_{1,1}-H_{2,2}}-e^{H_{3,3}-H_{2,2}}\right)
\nonumber\\
\partial_{t}S_{0}+\partial_{t}H_{2,3}  &  =-K_{1}e^{H_{3,3}-H_{2,3}%
}-2b\nonumber\\
\partial_{t}S_{0}+\partial_{t}H_{3,3}  &  =\frac{2b}{3}e^{H_{1,1}-H_{3,3}%
}-\frac{2b}{3}\left(  2-e^{H_{2,2}-H_{3,3}}\right)  . \label{U1E4}%
\end{align}

We first derive information about the asymptotics of the variables not
contained in the cycle (\ref{U1E1}). The equations for the variables
$H_{1,2},\ H_{2,2},\ H_{2,3}$ yield, neglecting lower order terms,
\begin{align*}
\partial_{t}S_{0}  &  =-K_{3}e^{H_{2,2}-H_{1,2}}+tK_{1}K_{3}e^{H_{2,3}%
-H_{1,2}}-K_{1}e^{H_{1,3}-H_{1,2}}\\
\partial_{t}S_{0}  &  =-2K_{1}e^{H_{2,3}-H_{2,2}}+\frac{2b}{3}e^{H_{1,1}%
-H_{2,2}}+\frac{2b}{3}e^{H_{3,3}-H_{2,2}}\\
\partial_{t}S_{0}  &  =-K_{1}e^{H_{3,3}-H_{2,3}}.
\end{align*}

We can eliminate from these equations the variables $H_{1,3},\ H_{3,3}$ using
(\ref{U1E3}) whence, after neglecting small order terms and using that
$\partial_{t}S_{0}$ scales like $t^{\frac{2}{3}}$%
\begin{align*}
\partial_{t}S_{0}  &  =-K_{3}e^{H_{2,2}-H_{1,2}}+tK_{1}K_{3}e^{H_{2,3}%
-H_{1,2}}-\frac{\partial_{t}S_{0}}{2K_{3}t}e^{H_{1,1}-H_{1,2}}\\
\partial_{t}S_{0}  &  =-2K_{1}e^{H_{2,3}-H_{2,2}}+\frac{2b}{3}e^{H_{1,1}%
-H_{2,2}}\\
1  &  =-\frac{1}{2tK_{3}}e^{H_{1,1}-H_{2,3}}.
\end{align*}

We can then write all the differences of functions $H_{j,k}$ in terms of
differences $\left(  H_{1,2}-H_{1,1}\right),$  $\left(  H_{2,2}-H_{1,1}%
\right)  $ and $\left(  H_{2,3}-H_{1,1}\right)  .$ Then, neglecting small
terms as $t\rightarrow\infty$, we obtain
\begin{align*}
\partial_{t}S_{0}  &  =-K_{3}e^{H_{1,1}-H_{1,2}}e^{H_{2,2}-H_{1,1}}%
-\frac{K_{1}}{2}e^{H_{1,1}-H_{1,2}}\\
\ \ e^{H_{1,1}-H_{2,2}}  &  =\frac{3}{2b}\partial_{t}S_{0}\ \ ,\ \ e^{H_{1,1}%
-H_{2,3}}=-2K_{3}t
\end{align*}
whence, after some computations,
\begin{equation}
e^{H_{1,1}-H_{2,3}}=-2K_{3}t\ \ ,\ \ e^{H_{1,1}-H_{2,2}}=\frac{3}{2b}%
\partial_{t}S_{0}\ \ ,\ \ e^{H_{1,1}-H_{1,2}}=-\frac{3}{2bK_{3}}\left(
1+\frac{3K_{1}}{4b}\right)  \left(  \partial_{t}S_{0}\right)  ^{2}.
\label{U1E2}%
\end{equation}

Notice that the three functions $e^{H_{2,3}},\ e^{H_{2,2}},\ e^{H_{1,2}}$ are
smaller than $e^{H_{1,1}}$ as $t\rightarrow\infty.$ Notice that some of these
functions can be negative, and therefore the corresponding function $H_{j,k}$
would have an additive complex factor $i\pi.$

We now compute the asymptotics of the functions $H_{1,1},\ H_{1,3},\ H_{3,3}.
$ Using the corresponding equations for the derivatives of these variables in
(\ref{U1E4}) and eliminating the variables $H_{1,2},\ H_{2,2},\ H_{2,3}$ using
(\ref{U1E2}) and neglecting small terms we obtain
\begin{align*}
\partial_{t}S_{0}+\partial_{t}H_{1,1}  &  =2tK_{1}K_{3}e^{H_{1,3}-H_{1,1}%
}+\frac{2b}{3}e^{H_{3,3}-H_{1,1}}-K_{2}e^{H_{1,3}-H_{1,1}}-\frac{4b}{3}\\
\partial_{t}S_{0}+\partial_{t}H_{1,3}  &  =tK_{1}K_{3}e^{H_{3,3}-H_{1,3}%
}-K_{2}e^{H_{3,3}-H_{1,3}}+\frac{1}{2t}e^{H_{1,1}-H_{1,3}}-2b\\
\partial_{t}S_{0}+\partial_{t}H_{3,3}  &  =\frac{2b}{3}e^{H_{1,1}-H_{3,3}%
}-\frac{4b}{3}+\frac{4b^{2}}{9}\frac{1}{\partial_{t}S_{0}}e^{H_{1,1}-H_{3,3}}.
\end{align*}

Using (\ref{U1E3}) to approximate the corrective exponential terms in these
equations we obtain
\begin{align*}
\partial_{t}S_{0}+\partial_{t}H_{1,1}  &  =2tK_{1}K_{3}e^{H_{1,3}-H_{1,1}%
}-\frac{4b}{3}\\
\partial_{t}S_{0}+\partial_{t}H_{1,3}  &  =tK_{1}K_{3}e^{H_{3,3}-H_{1,3}}-2b\\
\partial_{t}S_{0}+\partial_{t}H_{3,3}  &  =\frac{2b}{3}e^{H_{1,1}-H_{3,3}%
}-\frac{2b}{3}.
\end{align*}

Multiplying these equations, using $\left\vert \partial_{t}H_{j,k}\right\vert
\ll\partial_{t}S_{0}$ and keeping the leading order terms (using again
(\ref{U1E3}) to approximate the exponential terms on the right-hand side) we get
\[
\frac{\partial_{t}H_{1,1}}{\partial_{t}S_{0}}+\frac{\partial_{t}H_{1,3}%
}{\partial_{t}S_{0}}+\frac{\partial_{t}H_{3,3}}{\partial_{t}S_{0}}\sim-\left(
\frac{4b}{3}+2b+\frac{2b}{3}\right)  \frac{1}{\partial_{t}S_{0}}=-\frac
{4b}{\partial_{t}S_{0}}%
\]
whence, taking into account also that due to (\ref{U1E3}) we have
$H_{1,1}\sim H_{1,3}\sim H_{3,3}$ as $t\rightarrow\infty,$ it follows, to the
leading order
\[
H_{1,1}\sim H_{1,3}\sim H_{3,3}\sim-\frac{4b}{3}t\ \ \text{as\ \ }%
t\rightarrow\infty.
\]
We have thus obtained
\begin{equation}
S\sim\frac{3}{5}\left(  \frac{4b}{3}\right)  ^{\frac{1}{3}}\left(  K_{1}%
K_{3}\right)  ^{\frac{2}{3}}t^{\frac{5}{3}}-\frac{4b}{3}t\quad \text{as\ \ }%
t\rightarrow\infty. \label{U1E5}%
\end{equation}

\bigskip

\subsubsection{The asymptotics of $M_{j,k}$ are not compatible with the
self-similar behavior for the velocity distribution}

We now analyse if the asymptotics (\ref{U1E5}) is consistent with some of the
self-similar behaviors described in \cite{JNV1}, Section 5. The asymptotics
(\ref{U1E5}) for the tensor of second moments $M_{j,k},$ combined with the
mass conservation property, suggests the following long time behavior for the
solutions of (\ref{T6E9})
\begin{equation}
g\left(  t,w\right)  =\frac{1}{\left(  \lambda\left(  t\right)  \right)  ^{3}
}\Phi\left(  \xi\right)  \ ,\ \ \xi=\frac{w}{\lambda\left(  t\right)  }
\label{U1E6}%
\end{equation}
with
\[
\log\left(  \lambda\left(  t\right)  \right)  \sim\frac{3}{10}\left(
\frac{4b}{3}\right)  ^{\frac{1}{3}}\left(  K_{1}K_{3}\right)  ^{\frac{2}{3}%
}t^{\frac{5}{3}}
\quad \text{as\ \ }t\rightarrow\infty.
\]

Plugging (\ref{U1E6}) into (\ref{T6E9}) and keeping the leading order terms we
obtain the following equation for $\Phi$%
\[
-\frac{\partial_{t}\lambda}{\lambda}\Phi-\left[  K_{3}\xi_{2}+\left(
K_{2}-tK_{1}K_{3}\right)  \xi_{3}\right]  \partial_{\xi_{1}}\Phi-K_{1}\xi
_{3}\partial_{\xi_{2}}\Phi=\mathbb{C}\Phi\left(  \xi\right)  .
\]

Since $\frac{\partial_{t}\lambda}{\lambda}\sim\frac{1}{2}\left(  \frac{4b}%
{3}\right)  ^{\frac{1}{3}}\left(  K_{1}K_{3}\right)  ^{\frac{2}{3}}t^{\frac
{2}{3}}$ as $t\rightarrow\infty$ it is not possible to have a balance between
the terms $\frac{\partial_{t}\lambda}{\lambda}\Phi\left(  \xi\right)  $ and
$\left(  K_{1}K_{3}t\right)  \xi_{3}\partial_{\xi_{1}}\Phi$ unless
$\partial_{\xi_{1}}\Phi\rightarrow0$ as $t\rightarrow\infty.$ 
In that case $\Phi$ would be approximately independent on $\xi_1$ at least for a large set of values and this would be incompatible with $g$ having finite mass. Therefore, the
long time asymptotics of the solutions of (\ref{T6E9}) cannot be described by
a self-similar velocity distribution at least with the simple structure in
(\ref{U1E6}).

\bigskip


\subsection{A simple model with the hyperbolic terms much larger than the collision terms but yielding infinitely many collisions for long times \label{sec:domhyper}}

\subsubsection{Description of the model}

We have seen in some of the previous examples, that for some choices of the
matrices $L\left(  t\right)  $ and some
homogeneities of the kernels $B$ the collision term becomes much smaller than
the hyperbolic terms associated to the term $L\left(  t\right)  .$ However,
these collisions might yield huge deformations in the particle distribution.
In this subsection we consider a simplified model inspired by the dynamics of
the homoenergetic flows in the case of Simple Shear (cf. (\ref{T1E5})). We
have seen in Subsection 5.1 in \cite{JNV1} that in this case, if the
homogeneity $\gamma=0$, there are self-similar solutions for the
particle velocities. If $\gamma>0$ we have obtained in \cite{JNV2} a long time asymptotics described by a Maxwellian
distribution with temperature increasing in time as a power law. In Subsection \ref{SimpShearFrColl} we have seen that for
$\gamma<-1$ the average velocity of the particles increases due to the shear,
but therefore the collisions become so small that they yield a negligible
effect as $t\rightarrow\infty.$ If $\gamma\in\left[  -1,0\right)  $ the
description of the particle distribution seems more involved.

\bigskip

The main difficulty to describe the distribution of particles in the case of
Simple Shear and $\gamma\in\left[  -1,0\right)  $ is the following. The shear
is the dominant effect and it tends to yield a very elongated particle
distribution, analogous to the one obtained for the case $\gamma<-1$ (cf.
(\ref{A3E3}), (\ref{A3E4})). However, in the case $\gamma<-1$ the collision
rate decreases very quickly as $t\rightarrow\infty$ and a given particle
eventually does not experience any collision for long times. On the contrary,
if $\gamma\in\left[  -1,0\right)  $ the collision rate becomes small as
$t\rightarrow\infty,$ but each particle experiences infinitely many collisions
for long times, although increasingly spaced in time. The difficulty is that
the collision rule (\ref{CM1}), (\ref{CM2}) implies that the collision between
two particles whose velocities are in a distribution much elongated along the
axis $e_{1}$ get their directions deflected to an essentially arbitrary
direction (see Figure \ref{fig:ShearColl}). Therefore, in spite of the fact that the collisions preserve the
energy of the particles, the component $w_{2}$ of the particle velocities
increases in a significant way for most of the collisions. As a consequence
the increase of the velocities in the subsequent evolution by means of the
shear term becomes much larger than before the collision.

We now introduce a simple model for the evolution of a particle system under
the combined effect of shear and collisions containing some of the main
properties of the evolution described above. The resulting model is simpler
than the Boltzmann equation and it is possible to derive asymptotic formulas
for the behavior of its solutions.

\bigskip

We will assume that the particles of a system can be characterized by two real
variables, namely $\zeta>1$ and $\rho>0.$ We can think on $\zeta$ as the
component $\frac{w_{1}}{|w|}$ of the velocity in the Simple Shear case and
$\rho$ as the absolute value of the velocity $\left\vert w\right\vert .$ We
assume that between collisions $\zeta$ increases at a constant rate. Notice
that we can think also on $\zeta$ (more precisely $\left(  \zeta-1\right)  $)
as the time between collisions. On the other hand, we will assume that at the
collision times the particle jumps to a new value of $\rho,$ denoted as
$\tilde{\rho}$ and given by $\tilde{\rho}=\rho\zeta.$ The new value of $\zeta$
after the collision is reset to $\zeta=1.$ The collisions take place with the
rate $\varepsilon$ which typically will be assumed to be time dependent.

The particle distribution $f=f\left(  t,\rho,\zeta \right)  $ is then given by
\begin{align}
\partial_{t}f+\partial_{\zeta}f  &  =-\varepsilon\left(  t\right)  f\ \ ,\ \ \zeta
>1,\ \ \rho>0\ \ ,\ \ t>0\label{A1E4}\\
f\left(  t,\rho,1\right)   &  =\varepsilon\left(  t\right) \int_{1}^{\infty}f\left(
t,\frac{\rho}{\zeta},\zeta\right)  \frac{d\zeta}{\zeta}\,. \label{A1E5}%
\end{align}

We will assume that the initial particle distribution is
\begin{equation}
f\left(  0,\rho,\zeta\right)  =f_{0}\left(  \rho\right)  \delta\left(
\zeta-1\right)  . \label{A1E6}%
\end{equation}

Notice that in this model we are implicitly assuming that the absolute value
of the velocity for a particle characterized by the variables $\left(
\rho,\zeta\right)  $ is $\rho\zeta.$ We will study the dynamics of the model
(\ref{A1E4})-(\ref{A1E6}), in general with a time dependent $\varepsilon.$
However, we can also obtain a nonlinear version of (\ref{A1E4})-(\ref{A1E6})
in order to mimick the property of the Boltzmann equation according to which
for negative homogeneities of the kernel $B$ the collision rate decreases for
large particle velocities. In such nonlinear version of the model we consider
\begin{align}
\partial_{t}f+\partial_{\zeta}f  &  =-\varepsilon(t) f\ \ ,\ \ \zeta
>1,\ \ \rho>0\ \ ,\ \ t>0\label{A1E4_bis}\\
f\left(  t,\rho,1\right)   &  =\varepsilon(t)\int_{1}^{\infty}f\left(
t,\frac{\rho}{\zeta},\zeta\right)  \frac{d\zeta}{\zeta}\,. \label{A1E5_bis}%
\end{align}
with
\begin{equation}
\varepsilon\left(  t\right)  =\int_{0}^{\infty}d\rho\int_{1}^{\infty}%
d\zeta\ \frac{f\left(  t,\rho, \zeta \right)  }{\rho^a \zeta^{a}}\,,\quad a\in\left(
0,1\right)  . \label{A1E7}%
\end{equation}

The case $a\in\left(  0,1\right)  $ plays a role analogous to the homogeneity
of the kernel $\gamma\in\left(  -1,0\right)  $ in the case of Simple Shear for
the Boltzmann equation.

A difference between the models (\ref{A1E4})-(\ref{A1E6}), (\ref{A1E4_bis}%
)-(\ref{A1E7}) and the dynamics of the particles for homoenergetic solutions
of Boltzmann equation in the case of simple shear is that the particles can
move only in the direction of increasing $\zeta$ while in the Boltzmann case
$w_{1}$ can be increasing or decreasing. Notice that the smallness of
$\varepsilon$ (and therefore the long free flights between collisions) makes
reasonable to assume that the particles jump after each collision to
$\zeta=1,$ because 
we assumed that the jump takes place within a radius
of order $\tilde{\rho},$ since the length of the flight immediately later is
typically much larger than $\tilde{\rho}.$

It is readily seen that the solutions of (\ref{A1E4})-(\ref{A1E6}) (or
(\ref{A1E4_bis})-(\ref{A1E7})) satisfy
\begin{equation}
\partial_{t}\left(  \int_{0}^{\infty}d\rho\int_{1}^{\infty}d\zeta\  f\left(
t,\rho,\zeta \right)  \right)  =0 . \label{W3E2}%
\end{equation}

It is worth to notice that the Jacobian $\frac{d\zeta}{\zeta}$ in (\ref{A1E5})
is due to the multiplicative structure of the jumps. This Jacobian plays a
crucial role in the derivation of the mass conservation property (\ref{W3E2}).

The models (\ref{A1E4})-(\ref{A1E6}) with $\varepsilon\left(  t\right)
\rightarrow0$ as $t\rightarrow\infty$ or (\ref{A1E4_bis})-(\ref{A1E7}) have
several analogies with the homoenergetic flows for the Boltzmann equation. The
most relevant one is the existence of large particle flights which increase in
a significant manner the energy of the particle, followed by rare collisions
which transport the particle to a new state where the increase of energy due
to long flights is much larger than before. We will describe some results
concerning the asymptotics of the solutions of the models (\ref{A1E4}%
)-(\ref{A1E6}) and (\ref{A1E4_bis})-(\ref{A1E7}), which perhaps could shed some
light about the type of behaviors arising in the homoenergetic flows of the
Boltzmann equation in the cases in which the rate associated to the collision
terms tends to zero but it is nonintegrable.

\bigskip

\subsubsection{Reformulation of the model in an equivalent set of variables}

\bigskip

We now reformulate the models (\ref{A1E4})-(\ref{A1E6}) and (\ref{A1E4_bis}%
)-(\ref{A1E7}) in an alternative form which will make simpler to study their
long time asymptotics. We define $G\left(  t,X,Z\right)  $ by means of
\begin{equation}
f\left(  t,\rho,\zeta \right)  =\exp\left(  -\int_{0}^{t}\varepsilon\left(
\tau\right)  d\tau\right)  G\left( t, X,Z\right)  \ \ ,\ \rho=e^{X}%
\ \ ,\ \ \zeta=e^{Z} . \label{W3E3}%
\end{equation}

Then (\ref{A1E4})-(\ref{A1E6}) become%
\begin{align}
\partial_{t}G+e^{-Z}\partial_{Z}G  &  =0\ \ ,\ \ Z>0\ ,\ X\in\mathbb{R}%
\label{A2E1}\\
G\left(  t,X,0\right)   &  =\varepsilon\left(  t\right)  \int_{0}^{\infty
}G\left(  t,X-Z,Z\right)  dZ\label{A2E2}\\
G\left(  0, X,Z\right)   &  =G_{0}\left(  X\right)  \delta\left(  Z\right)  .
\label{A2E3}%
\end{align}

We can reformulate (\ref{A2E1})-(\ref{A2E3}) as an integral equation for the
function $\Phi\left(  t,X\right) =G\left( t, X,0\right)  .$ Integrating by
characteristics (\ref{A2E1}),\ (\ref{A2E3}) we obtain
\begin{align}
G\left(  t,X,Z\right)   &  =\frac{G_{0}\left(  X\right)  }{1+t}\delta\left(
Z-\log\left(  1+t\right)  \right)  +G\left( t+1-e^{Z}, X,0\right) \nonumber\\
&  =\frac{G_{0}\left(  X\right)  }{1+t}\delta\left(  Z-\log\left(  1+t\right)
\right)  +\Phi\left(  t+1-e^{Z},X\right)  . \label{A2E3a}%
\end{align}

Using this formula in (\ref{A2E2}) we obtain
\[
\Phi\left(  t,X\right) =\frac{\varepsilon\left(  t\right)  }{1+t}\int
_{0}^{\infty}G_{0}\left(  X-Z\right)  \delta\left(  Z-\log\left(  1+t\right)
\right)  dZ+\varepsilon\left(  t\right)  \int_{0}^{\log\left(  1+t\right)
}\Phi\left(  t+1-e^{Z},X-Z\right)  dZ
\]
and using the change of variables $e^{Z}-1=\xi$ we then obtain the integral equation%

\begin{equation}
\Phi\left(  t,X\right) =\frac{\varepsilon\left(  t\right)  }{1+t}G_{0}\left(
X-\log\left(  1+t\right)  \right)  +\varepsilon\left(  t\right)  \int_{0}%
^{t}\Phi\left( t-\xi, X-\log\left(  1+\xi\right)  \right)  \frac{d\xi}{1+\xi}.
\label{W3E4}%
\end{equation}

In the case in which $\varepsilon\left(  t\right)  $ is given by (\ref{A1E7})
we obtain, using (\ref{W3E3})
\[
\varepsilon\left(  t\right)  \exp\left(  \int_{0}^{t}\varepsilon\left(
\tau\right)  d\tau\right)  =\int_{-\infty}^{\infty}e^{\left(  1-a\right)
X}dX\int_{0}^{\infty}e^{\left(  1-a\right)  Z}G\left(  t,X,Z\right)  dZ
\]
and using (\ref{A2E3a}) we then obtain
\begin{equation}
\varepsilon\left(  t\right)  \exp\left(  \int_{0}^{t}\varepsilon\left(
\tau\right)  d\tau\right)  =\frac{C_{0}}{\left(  1+t\right)  ^{a}}%
+\int_{-\infty}^{\infty}e^{\left(  1-a\right)  X}dX\int_{0}^{t}\Phi\left(
t-\xi,X\right)  \frac{d\xi}{\left(  1+\xi\right)  ^{a}}\ \label{W3E5}%
\end{equation}
where
\begin{equation}
C_{0}=\int_{-\infty}^{\infty}G_{0}\left(  X\right)  e^{\left(  1-a\right)
X}dX. \label{W3E5a}%
\end{equation}

We have then reduced the models (\ref{A1E4})-(\ref{A1E6}) and (\ref{A1E4_bis}%
)-(\ref{A1E7}) to the models (\ref{W3E4}) and (\ref{W3E4})-(\ref{W3E5a})
respectively. We now study the asymptotics of the solutions of these models.
We begin studying (\ref{W3E4}) with constant $\varepsilon.$

\bigskip

\subsubsection{The model (\ref{W3E4}) with constant $\varepsilon$
\label{ConstEps}}

The equation (\ref{W3E4}) can be explicitly solved using Fourier and Laplace
transforms if $\varepsilon\left(  t\right)  =\varepsilon$ is a constant, i.e.
\begin{equation}
\Phi\left(  t,X\right) =\frac{\varepsilon}{1+t}G_{0}\left(  X-\log\left(
1+t\right)  \right)  +\varepsilon\int_{0}^{t}\Phi\left(  t-\xi, X-\log\left(
1+\xi\right)\right)  \frac{d\xi}{1+\xi},\ \ X\in\mathbb{R}\ ,\ t\geq0.
\label{W3E6}%
\end{equation}

Actually we need to consider a more general class of problems. Given any
$\beta\in\mathbb{R}$ we define functions%
\begin{equation}
\Psi\left(  t,X\right)  =\Psi_{\beta}\left( t, X\right)  =\Phi\left(
t,X\right)  e^{\beta X}. \label{W3E6a}%
\end{equation}

Then, formally, the functions $\Psi$ solve the equations
\begin{equation}
\Psi\left(  t,X\right)  =\frac{\varepsilon}{\left(  1+t\right)  ^{1-\beta}%
}H_{0}\left(  X-\log\left(  1+t\right)  \right)  +\varepsilon\int_{0}^{t}%
\Psi\left(  t-\xi, X-\log\left(  1+\xi\right) \right)  \frac{d\xi}{\left(
1+\xi\right)  ^{1-\beta}},\ \ \label{W3E8}%
\end{equation}
for $X\in\mathbb{R}$ and $t\geq0 $. Here
\begin{equation}
H_{0}\left(  X\right)  =H_{0,\beta}\left(  X\right)  =G_{0}\left(  X\right)
e^{\beta X} . \label{W3E8a}%
\end{equation}

We will assume that $G_{0}$ decreases fast enough as $\left\vert X\right\vert
\rightarrow\infty$ in order to guarantee the convergence of all the integrals
appearing later. We use the following form of the Fourier transform in $X$%
\begin{equation}
\psi\left( t, k\right)  =\frac{1}{\sqrt{2\pi}}\int_{-\infty}^{\infty}%
\Psi\left(  t,X\right)  e^{-ikX}dX. \label{W3E7}%
\end{equation}

Notice that the Fourier transform $\psi\left(  t,k\right)  $ is defined if
$\Psi\left(  t,X \right)  $ does not increase exponentially as $\left\vert
X\right\vert \rightarrow\infty.$ It is well known that the Fourier transform
can be defined in the class of tempered distributions. We will not make
precise the spaces in which the functions $\Psi$ are, but this could be made
easily using the properties of Fourier and Laplace transforms in distribution
spaces. (See for instance \cite{DK}).

Taking formally the Fourier transform in $X$ of (\ref{W3E8}) we obtain
\begin{equation}
\psi\left(  t,k\right)  =\frac{\varepsilon}{\left(  1+t\right)  ^{1-\beta+ik}%
}h_{0}\left(  k\right)  +\varepsilon\int_{0}^{t}\frac{\psi\left(t-\xi,
k\right)  d\xi}{\left(  1+\xi\right)  ^{1-\beta+ik}} \label{W3E9}%
\end{equation}
where $h_{0}$ is the Fourier transform of $H_{0}.$ In order to solve
(\ref{W3E9}) we take the Laplace transform in time. We define the Laplace
transform as
\begin{equation}
\tilde{\psi}\left(  z,k\right)  =\int_{0}^{\infty}\psi\left(  t,k\right)
e^{-zt}dt . \label{W4E1}%
\end{equation}

Then using (\ref{W3E9}) and the properties of the Laplace transform of a
convolution we obtain
\[
\tilde{\psi}\left(  z,k\right)  \left[  1-\varepsilon\Lambda\left(
z,k\right)  \right]  =\varepsilon h_{0}\left(  k\right)  \Lambda\left(
z,k\right)
\]
where
\begin{equation}
\Lambda\left(  z,k\right)  =\Lambda_{\beta}\left( z,k\right)  =\int
_{0}^{\infty}\frac{e^{-zt}dt}{\left(  1+t\right)  ^{1-\beta+ik}}\, .
\label{W4E2}%
\end{equation}

Therefore, as long as $\left[  1-\varepsilon\Lambda\left(  z,k\right)
\right]  \neq0$ we obtain
\begin{equation}
\tilde{\psi}\left(  z,k\right)  =\frac{\varepsilon h_{0}\left(  k\right)
\Lambda\left(  z,k\right)  }{1-\varepsilon\Lambda\left(  z,k\right)  } .
\label{W4E3}%
\end{equation}

The Fourier modes with a given $k\in\mathbb{R}^{3}$ increase exponentially as
$\exp\left(  z_{0}\left(  k;\varepsilon\right)  t\right)  $ where
$z_{0}\left(  k;\varepsilon\right)  $ is the solution $z$ of the equation
\begin{equation}
1-\varepsilon\Lambda\left(  z,k\right)  =0 \label{W4E3a}%
\end{equation}
with the largest real part. We then need to understand the roots of the
equation
\begin{equation}
\frac{1}{\varepsilon}=\int_{0}^{\infty}\frac{e^{-zt}dt}{\left(  1+t\right)
^{1-\beta+ik}} . \label{W4E4}%
\end{equation}

For any $\beta\geq0,$ there exists a unique root $z_{0}\left(  0;\varepsilon
\right)  \in\mathbb{R}_{+}$ of (\ref{W4E4}) for $k=0.$ This follows from the
fact that the function
\begin{equation}
z\rightarrow\Lambda\left(  z,0\right)  =\int_{0}^{\infty}\frac{e^{-zt}%
dt}{\left(  1+t\right)  ^{1-\beta}}\ \ ,\ \ z\in\mathbb{R}_{+}\ \label{W4E5}%
\end{equation}
is decreasing in $z$ and it converges to infinity as $z\rightarrow0^{+}.$
Moreover, given any other root $z_{0}\left(  k;\varepsilon\right)  $ of
(\ref{W4E4}) with $k\neq0$ we have $\operatorname{Re}\left(  z_{0}\left(
k;\varepsilon\right)  \right)  <z_{0}\left(  0;\varepsilon\right)  .$ Indeed,
if $k\neq0$ we can write
\[
\int_{0}^{\infty}\frac{e^{-\operatorname{Re}\left(  z_{0}\left(
0;\varepsilon\right)  \right)  t}dt}{\left(  1+t\right)  ^{1-\beta}}=\frac
{1}{\varepsilon}=\int_{0}^{\infty}\frac{e^{-zt}dt}{\left(  1+t\right)
^{1-\beta+ik}}<\int_{0}^{\infty}\frac{e^{-\operatorname{Re}\left(
z_{0}\left(  k;\varepsilon\right)  \right)  t}dt}{\left(  1+t\right)
^{1-\beta}}%
\]
and using the fact that the function defined in (\ref{W4E5}) is decreasing we
obtain $\operatorname{Re}\left(  z_{0}\left(  0;\varepsilon\right)  \right)
>\operatorname{Re}\left(  z_{0}\left(  k;\varepsilon\right)  \right)  .$

Moreover, we can compute the asymptotics of $z_{0}\left(  0;\varepsilon
\right)  $ using the asymptotics
\[
\int_{0}^{\infty}\frac{e^{-zt}dt}{\left(  1+t\right)  ^{1-\beta}}=\frac
{1}{z^{\beta}}\int_{0}^{\infty}\frac{e^{-\zeta}d\zeta}{\left(  z+\zeta\right)
^{1-\beta}}\sim\frac{\Gamma\left(  \beta\right)  }{z^{\beta}}\text{
as\ }z\rightarrow0^{+}%
\]
whence
\begin{equation}
z_{0}\left(  0;\varepsilon\right)  \sim\left(  \Gamma\left(  \beta\right)
\varepsilon\right)  ^{\frac{1}{\beta}}\ \ \text{as\ }\varepsilon
\rightarrow0\,\ \text{if }\beta>0\ . \label{W4E6}%
\end{equation}

If $\beta=0,$ we obtain that $z_{0}\left(  0;\varepsilon\right)  $ is
exponentially small, due to the logarithmic divergence of the integral in
(\ref{W4E5}) as $z\rightarrow0^{+},$ but we will not need the detailed
analysis of $\beta$ in such a case.

Notice that using (\ref{W4E3}) and (\ref{W4E6}) as well as the inversion
formula for the Laplace transform we obtain, for $k=0$, the following approximation for
small $\varepsilon$%
\[
\psi\left(  t,0\right)  \sim\frac{h_{0}\left(  0\right)  }{B\varepsilon}%
\exp\left(  \left(  \Gamma\left(  \beta\right)  \varepsilon\right)  ^{\frac
{1}{\beta}}t\right)\quad   \text{ as }\quad t\rightarrow\infty
\]
where
\begin{equation}
B=-\frac{\partial\Lambda\left(  z_{0}\left(  0;\varepsilon\right) ,0 \right)
}{\partial z}=\int_{0}^{\infty}\frac{e^{-z_{0}\left(  0;\varepsilon\right)
t}t}{\left(  1+t\right)  ^{1-\beta}}dt\sim\frac{\Gamma\left(  \beta+1\right)
}{\left(  z_{0}\left(  0;\varepsilon\right)  \right)  ^{1+\beta}}=\frac{\beta
}{\left(  \Gamma\left(  \beta\right)  \right)  ^{\frac{1}{\beta}}}\frac
{1}{\varepsilon^{1+\frac{1}{\beta}}}\;\ \text{ as }\; \varepsilon\rightarrow0. \label{eq:formulaB}
\end{equation}

Then
\[
\psi\left(  t,0\right)  \sim\frac{\left(  \Gamma\left(  \beta\right)
\varepsilon\right)  ^{\frac{1}{\beta}}}{\beta}h_{0}\left(  0\right)
\exp\left(  \left(  \Gamma\left(  \beta\right)  \varepsilon\right)  ^{\frac
{1}{\beta}}t\right)  \text{ as }t\rightarrow\infty.
\]

Using the definitions of the Fourier transform (cf. (\ref{W3E7})) and of the functions $H_{0},\ \Psi$ (cf. (\ref{W3E6a}), (\ref{W3E8a})) we obtain
\begin{equation}
\int_{-\infty}^{\infty}\Phi\left(  t,X\right) e^{\beta X}dX\sim\frac{\left(
\Gamma\left(  \beta\right)  \varepsilon\right)  ^{\frac{1}{\beta}}}{\beta}%
\exp\left(  \left(  \Gamma\left(  \beta\right)  \varepsilon\right)  ^{\frac
{1}{\beta}}t\right)  \int_{-\infty}^{\infty}G_{0}\left(  X\right)  e^{\beta
X}dX\text{ as }t\rightarrow\infty. \label{W4E7}%
\end{equation}

The formula (\ref{W4E7}) provides a large amount of information about the
distribution of mass of the function $\Phi\left(  t,X\right) $ as
$t\rightarrow\infty.$ It is interesting to remark that the integrals
$\int_{-\infty}^{\infty}\Phi e^{\beta X}dX$ increase at a different rate for
different values of $\beta.$ Due to the changes of variables in (\ref{W3E3})
this would imply that different moments of $f$ increase at a different rate.
As indicated in previous subsections this is incompatible with a self-similar
behavior for $f.$

Actually, it is possible to obtain some additional information about the
transport of mass towards $X\rightarrow\infty$ for the solutions of
(\ref{W3E8}) deriving the asymptotics of $z_{0}\left(  k;\varepsilon\right)  $
as $k\rightarrow 0$ by means of (\ref{W4E4}). This yields
\begin{equation}
z_{0}\left(  k;\varepsilon\right)  \sim z_{0}\left(  0;\varepsilon\right)
+A^{\eps}_{1}ik-A^{\eps}_{2}k^{2}\text{ as }k\rightarrow0 \label{W4E8}%
\end{equation}
for suitable real functions $A^{\eps}_{1},\ A^{\eps}_{2}>0$ depending on $\varepsilon$ and
$\beta.$ Using the formula for the inversion of Fourier as well as
(\ref{W4E8}) it is possible to obtain an expansion for $\Phi\left(
X,t\right)  $ with the form
\begin{equation}
\Phi\left(  t,X\right) \sim \frac{\varepsilon h_{0}\left(  0\right)  \Lambda\left(  z_{0}\left(
0;\varepsilon\right)  ,0\right)  }{B_{\varepsilon}\left(  0\right)  }%
\frac{\exp\left(  z_{0}\left(  0;\varepsilon\right)  t\right)  }{\sqrt{A^{\eps}_{2}%
t} e^{\beta X}}Q_{\varepsilon}\left(  \frac{X+A^{\eps}_{1}t}{\sqrt{A^{\eps}_{2}t}}\right)
 \,\text{\ as
}t\rightarrow\infty, \label{eq:asympPhi}
\end{equation}
where 
\[
Q_{\varepsilon}\left(  \xi\right)  
=\frac{1}{\sqrt{2}}\exp\left(
-\frac{\xi^{2}}{4}\right).
\]
We want to give a more detailed expression of the formula \eqref{eq:asympPhi}. 

We will derive the asymptotics for $\Phi\left(  t,X\right)  $ using \eqref{W3E6a}, namely $\Psi\left(  t,X\right)  =\Phi\left(
t,X\right)  e^{\beta X}$. We
need to compute the asymptotics for $\Psi\left(  t,X\right)  .$ To this end we
invert the Laplace transform $\tilde{\psi}\left(  z,k\right)  $ in \eqref{W4E3}, i.e.~we compute
\begin{equation}
\psi\left(  t,k\right)  =\frac{1}{2\pi i}\int_{C}\frac{\varepsilon
h_{0}\left(  k\right)  \Lambda\left(  z,k\right)  }{1-\varepsilon
\Lambda\left(  z,k\right)  }e^{zt}dz\label{IntA1}.
\end{equation}
There is a zero of $\left(  1-\varepsilon\Lambda\left(  z,k\right)  \right)  $
at $z=z_{0}\left(  k;\varepsilon\right)$.  We assume that: 
\[
1-\varepsilon\Lambda\left(  z,k\right)  =B_{\varepsilon}\left(  k\right)
\left(  z-z_{0}\left(  k;\varepsilon\right)  \right)  \left[  1+o\left(
1\right)  \right] \quad \text{ as }\quad z\rightarrow z_{0}\left(  k;\varepsilon\right).
\]
Thus, we can compute the integral in (\ref{IntA1}) using residues. We observe that there would
be additional contributions to the integral, but they are smaller as
$t\rightarrow\infty.$ We then obtain the asymptotics:
\begin{equation}
\psi\left(  t,k\right)  \sim\frac{\varepsilon h_{0}\left(  k\right)
\Lambda\left(  z_{0}\left(  k;\varepsilon\right)  ,k\right)  }{B_{\varepsilon
}\left(  k\right)  }\exp\left(  z_{0}\left(  k;\varepsilon\right)  t\right)
\ \ \quad \text{as }\quad t\rightarrow\infty . \label{IntA2}
\end{equation}
The function $B\left(  k\right)  $ is obtained by means of the derivative of
$\Lambda\left(  z,k\right) $ (cf. \eqref{eq:formulaB} in the case $k=0$). 
The formula of $B\left(
k\right)  $ is:
\[
B_{\varepsilon}\left(  k\right)  =\int_{0}^{\infty}\frac{e^{-z_{0}\left(
k;\varepsilon\right)  t}}{\left(  1+t\right)  ^{1-\beta+ik}}dt .
\]

We are computing the asymptotics of the solutions as $k\rightarrow0.$ We use
the asymptotic formula \eqref{W4E8}. If we fix $\varepsilon$ and we take
$k\rightarrow0$ we obtain that the function $B_{\varepsilon}\left(  k\right)
$ is continuous in $k$ (for each $\varepsilon>0$ fixed) and we have
$\lim_{k\rightarrow0}B_{\varepsilon}\left(  k\right)  =B_{\varepsilon}\left(
0\right)  .$

\bigskip

Therefore, taking the asymptotics $k\rightarrow0$, we obtain the following
approximation for (\ref{IntA2}): 
\begin{equation}
\psi\left(  t,k\right)  \sim\frac{\varepsilon h_{0}\left(  0\right)
\Lambda\left(  z_{0}\left(  0;\varepsilon\right)  ,0\right)  }{B_{\varepsilon
}\left(  0\right)  }\exp\left(  \left[  z_{0}\left(  0;\varepsilon\right)
+A^{\eps}_{1}ik-A^{\eps}_{2}k^{2}\right]  t\right)  \label{IntA3}%
\end{equation}
where we use \eqref{W4E8} in the approximation of the exponent. We assume also that
$h_{0}$ is smooth and $h_{0}\left(  0\right)  >0.$

We can obtain now the inverse of the Fourier transform to derive the
asymptotics of $\Psi\left(  t,X\right)  $ and then $\Phi\left(  t,X\right)  $
using \eqref{W3E6a}. We have:%
\[
\Psi\left(  t,X\right)  =\frac{1}{\sqrt{2\pi}}\int_{-\infty}^{\infty}%
\psi\left(  t,k\right)  e^{ikX}dk.
\]

We then obtain, using (\ref{IntA3}), the asymptotics:%
\begin{align*}
\Psi\left(  t,X\right)    & \sim\frac{1}{\sqrt{2\pi}}\frac{\varepsilon
h_{0}\left(  0\right)  \Lambda\left(  z_{0}\left(  0;\varepsilon\right)
,0\right)  }{B_{\varepsilon}\left(  0\right)  }\int_{-\infty}^{\infty}%
\exp\left(  \left[  z_{0}\left(  0;\varepsilon\right)  +A^{\eps}_{1}ik-A^{\eps}_{2}%
k^{2}\right]  t\right)  e^{ikX}dk\\
& =\frac{1}{\sqrt{2\pi}}\frac{\varepsilon h_{0}\left(  0\right)
\Lambda\left(  z_{0}\left(  0;\varepsilon\right)  ,0\right)  }{B_{\varepsilon
}\left(  0\right)  }\exp\left(  z_{0}\left(  0;\varepsilon\right)  t\right)
\int_{-\infty}^{\infty}\exp\left(  -A^{\eps}_{2}k^{2}t\right)  e^{ik\left(
X+A^{\eps}_{1}t\right)  }dk
\end{align*}
and inverting the Fourier transform we get \eqref{eq:asympPhi}. 
Using this formula and \eqref{W3E6a} we obtain \eqref{W4E7}. 

Using these expansions for arbitrary values of $\beta$ it is possible to
derive a large amount of information about the asymptotics of $\Phi\left(
t,X\right)  $ in different regions of the plane $\left(  X,t\right)  $ as
$t\rightarrow\infty.$ However, we will not use this type of detailed
asymptotic formulas in this paper.

\subsubsection{The model (\ref{W3E4}) with slowly changing $\varepsilon$
\label{SlowEps}}

We now examine the equation (\ref{W3E4}) assuming that the function
$\varepsilon\left(  t\right)  $ changes slowly. By this we mean that
$\left\vert \partial_{t}\varepsilon\left(  t\right)  \right\vert
\ll\varepsilon\left(  t\right)  .$ A typical behavior for $\varepsilon\left(
t\right)  $ would be $\varepsilon\left(  t\right)  \sim\frac{A}{t}$ as
$t\rightarrow\infty$ for some constant $A>0.$ Suppose that for any $\beta
\geq0$ we define $\Psi$ is as in (\ref{W3E6a}). Actually, we are interested in
the asymptotics of 
$\int_{-\infty}^{\infty}\Psi\left(  t,X\right)
dX=\lambda\left(  t\right)  =\lambda_{\beta}\left(  t\right)  .$ Then, using
(\ref{W3E8}) we obtain
\begin{equation}
\lambda\left(  t\right)  =\frac{C_{\beta}\varepsilon\left(  t\right)
}{\left(  1+t\right)  ^{1-\beta}}+\varepsilon\left(  t\right)  \int_{0}%
^{t}\frac{\lambda\left(  t-\xi\right)  d\xi}{\left(  1+\xi\right)  ^{1-\beta}}
\label{W4E9}%
\end{equation}
where
\[
C_{\beta}=\int_{-\infty}^{\infty}H_{0}\left(  X\right)  dX.
\]

We look for solutions of (\ref{W4E9}) with the form
\[
\lambda\left(  t\right)  =\exp\left(  \int_{0}^{t}z\left(  \tau\right)
d\tau\right)
\]
for some suitable function $z\left(  \tau\right)  $ to be determined. Then
\begin{equation}
1=\frac{C_{\beta}\varepsilon\left(  t\right)  \exp\left(  -\int_{0}%
^{t}z\left(  \tau\right)  d\tau\right)  }{\left(  1+t\right)  ^{1-\beta}%
}+\varepsilon\left(  t\right)  \int_{0}^{t}\frac{\exp\left(  -\int_{t-\xi}%
^{t}z\left(  \tau\right)  d\tau\right)  d\xi}{\left(  1+\xi\right)  ^{1-\beta
}} . \label{W5E1}%
\end{equation}

The first term on the right can be expected to be exponentially small compared
with the second. On the other hand if $\varepsilon\left(  t\right)  $ changes
slowly in the form indicated above we expect $z\left(  \tau\right)  $ to be
approximately given by $z_{0}\left(  0;\varepsilon\left(  t\right)  \right)
.$ Indeed, if we assume that $z\left(  \tau\right)  $ changes slowly in $\tau$
and we neglect the first term on the right-hand side of (\ref{W5E1}) we obtain
the approximation
\begin{equation}
\frac{1}{\varepsilon\left(  t\right)  }=\int_{0}^{t}\frac{\exp\left(
-z\left(  t\right)  \xi\right)  }{\left(  1+\xi\right)  ^{1-\beta}}d\xi.
\label{W5E1a}%
\end{equation}

Suppose that $z\left(  t\right)  \rightarrow0$ as $t\rightarrow\infty.$ We can
then approximate the integral on the right-hand side of (\ref{W5E1a}) in the same manner
as (\ref{W4E4}), i.e. we write
\[
\int_{0}^{t}\frac{\exp\left(  -z\left(  t\right)  \xi\right)  }{\left(
1+\xi\right)  ^{1-\beta}}d\xi=\frac{1}{z\left(  t\right)  }\int_{0}^{\frac
{t}{z\left(  t\right)  }}\frac{\exp\left(  -y\right)  }{\left(  1+\frac
{y}{z\left(  t\right)  }\right)  ^{1-\beta}}dy\sim\frac{1}{\left(  z\left(
t\right)  \right)  ^{\beta}}\int_{0}^{\infty}\frac{\exp\left(  -y\right)
}{y^{1-\beta}}dy=\frac{\Gamma\left(  \beta\right)  }{\left(  z\left(
t\right)  \right)  ^{\beta}}%
\]
whence, arguing as in the derivation of (\ref{W4E6}),
\begin{equation}
z\left(  t\right)  \sim z_{0}\left(  0;\varepsilon\left(  t\right)  \right)
\sim\left(  \Gamma\left(  \beta\right)  \varepsilon\left(  t\right)  \right)
^{\frac{1}{\beta}}\ \ \text{as\ }\varepsilon\rightarrow0\,\ \text{if }\beta>0
. \label{W5E2}%
\end{equation}

In order to check the validity of the approximation we just need to check the
assumptions made in its derivation. The two assumptions made in the derivation
of (\ref{W5E2}) are the following ones:
\begin{equation}
\frac{C_{\beta}\varepsilon\left(  t\right)  }{\left(  1+t\right)  ^{1-\beta}%
}\ll\lambda\left(  t\right)  \ \ \text{as\ \ }t\rightarrow\infty\ \label{W5E3}%
\end{equation}
and%
\begin{equation}
\int_{0}^{t}\frac{\exp\left(  -\int_{t-\xi}^{t}z\left(  \tau\right)
d\tau\right)  d\xi}{\left(  1+\xi\right)  ^{1-\beta}}\sim\int_{0}^{t}%
\frac{\exp\left(  -z\left(  t\right)  \xi\right)  }{\left(  1+\xi\right)
^{1-\beta}}d\xi\text{ as }t\rightarrow\infty. \label{W5E4}%
\end{equation}

The approximation (\ref{W5E4}) holds, assuming that $z\left(  t\right)  $
behaves like (\ref{W5E2}) if $\varepsilon\left(  t\right)  \sim\frac
{A}{t}$ as $t\rightarrow\infty$ and $\beta>1.$ Indeed, in that case we have
\begin{align*}
&  \int_{0}^{t}\frac{\exp\left(  -\int_{t-\xi}^{t}z\left(  \tau\right)
d\tau\right)  }{\left(  1+\xi\right)  ^{1-\beta}}d\xi\\
&  \sim\int_{0}^{t}\frac{\exp\left(  -\left(  \Gamma\left(  \beta\right)
A\right)  ^{\frac{1}{\beta}}\int_{t-\xi}^{t}\tau^{-\frac{1}{\beta}}%
d\tau\right)  }{\left(  1+\xi\right)  ^{1-\beta}}d\xi\\
&  =\int_{0}^{t}\frac{\exp\left(  -\frac{\left(  \Gamma\left(  \beta\right)
A\right)  ^{\frac{1}{\beta}}\beta}{\beta-1}\left[  \left(  t\right)
^{-\frac{1}{\beta}+1}-\left(  t-\xi\right)  ^{-\frac{1}{\beta}+1}\right]
\right)  }{\left(  1+\xi\right)  ^{1-\beta}}d\xi\\
&  =t\int_{0}^{1}\frac{\exp\left(  -\frac{\left(  \Gamma\left(  \beta\right)
A\right)  ^{\frac{1}{\beta}}\beta}{\beta-1}\left(  t\right)  ^{-\frac{1}%
{\beta}+1}\left[  1-\left(  1-\eta\right)  ^{-\frac{1}{\beta}+1}\right]
\right)  }{\left(  1+t\eta\right)  ^{1-\beta}}d\eta
\end{align*}
and this integral can be approximated if $t\rightarrow\infty$, using the
Laplace method (cf.~\cite{BeOr}), by means of the right-hand side of (\ref{W5E4}). On the other
hand (\ref{W5E3}) holds in this case, since $\lambda\left(  t\right)  $ tends
exponentially to infinity.

\bigskip

Nevertheless, if $\beta<1$ and $\varepsilon\left(  t\right)  \sim\frac{A}{t}$
as $t\rightarrow\infty$ the solution (\ref{W5E2}) would not describe the
asymptotics of $\lambda\left(  t\right)  $ because the approximation
(\ref{W5E3}) would fail. In this case we will try a solution of (\ref{W5E1})
with the form
\begin{equation}
\lambda\left(  t\right)  =e^{J\left(  t\right)  } . \label{W6E2a}%
\end{equation}

We consider the case with $\varepsilon\left(  t\right)  \sim\frac{A}{t}.$ It
will turn out that in this case we will not be able to assume that
$\frac{C_{\beta}\varepsilon\left(  t\right)  }{\left(  1+t\right)  ^{1-\beta}%
}\ll\lambda\left(  t\right)  \ $as\ $t\rightarrow\infty$ (see (\ref{W5E3})).
We have and we would have the approximated problem
\begin{equation}
1=\frac{C_{\beta}\varepsilon\left(  t\right)  }{\left(  1+t\right)  ^{1-\beta
}\lambda\left(  t\right)  }+\frac{A}{t}\int_{0}^{t}\frac{\exp\left(  J\left(
t-\xi\right)  -J\left(  t\right)  \right)  }{\xi^{1-\beta}}d\xi. \label{W6E1}%
\end{equation}

We can obtain an approximate solution of (\ref{W6E1}) in the form
\begin{equation}
J\left(  t\right)  =-B\left[  \log\left(  t+1\right)  \right]  \label{W6E2}%
\end{equation}
for some suitable $B>0.$ Then (\ref{W6E1}) becomes for large $t$
\[
1=\frac{C_{\beta}\varepsilon\left(  t\right)  \left(  t+1\right)  ^{B}%
}{\left(  1+t\right)  ^{1-\beta}}+\frac{A\left(  t+1\right)  ^{B}}{t}\int
_{0}^{t}\frac{1}{\xi^{1-\beta}}\frac{1}{\left(  t-\xi+1\right)  ^{B}}d\xi.
\]

If $B>1$ we can approximate the right-hand side of this equation as%
\[
\frac{C_{\beta}A\left(  t+1\right)  ^{B}}{\left(  1+t\right)  ^{1-\beta}%
t}+\frac{A\left(  t+1\right)  ^{B}}{t}\frac{1}{t^{1-\beta}}\int_{0}^{t}%
\frac{1}{\left(  t-\xi+1\right)  ^{B}}d\xi=\frac{AC_{0}\left(  t+1\right)
^{B}}{t^{2-\beta}}%
\]
and then we obtain a possible solution of (\ref{W6E1}) if $B=2-\beta$ which is
larger than one, and therefore gives a consistent asymptotics. Notice,
however, that the determination of the multiplicative constants, in particular
$C_{0}$ requires a more careful analysis, because this quantity is really
determined by the values of $\lambda\left(  t\right)  $ with $t$ of order one.
A more careful examination of the argument shows that
\[
C_{0}\simeq C_{\beta}A+\int_{0}^{\infty}\lambda\left(  t\right)  dt.
\]

In any case, we just indicate at this point that the assumptions (\ref{W5E3}),
(\ref{W5E4}) which are at the basis of the approximation of the solution
$\lambda\left(  t\right)  $ by means of an adiabatic change of the eigenvalue
$z\left(  t\right)  $ cannot be given by granted for arbitrary values of
$\beta$ and $\varepsilon\left(  t\right)  $. In particular, a careful analysis
of the conditions (\ref{W5E3}), (\ref{W5E4}) is needed in each particular case.

\subsubsection{The model (\ref{W3E4})-(\ref{W3E5a})}

Finally we derive asymptotic formulas for the solutions $\varepsilon\left(
t\right)  $ of the problem (\ref{W3E4})-(\ref{W3E5a}). To this end we multiply
(\ref{W3E4}) by $e^{\left(  1-a\right)  X}$ and integrating in $X$ we obtain
the following equation for $\lambda\left(  t\right)  =\int_{-\infty}^{\infty
}e^{\left(  1-a\right)  X}\Phi\left(  t,X\right) dX$ (cf. also (\ref{W4E9}))%
\begin{equation}
\lambda\left(  t\right)  =\frac{C_{a}\varepsilon\left(  t\right)  }{\left(
1+t\right)  ^{a}}+\varepsilon\left(  t\right)  \int_{0}^{t}\frac
{\lambda\left(  t-\xi\right)  d\xi}{\left(  1+\xi\right)  ^{a}} \label{W5E5}%
\end{equation}
with
\[
C_{a}=\int_{-\infty}^{\infty}G_{0}\left(  X\right)  e^{\left(  1-a\right)
X}dX\text{ and }0<a<1.
\]

On the other hand (\ref{W3E5}) becomes
\begin{equation}
\varepsilon\left(  t\right)  \exp\left(  \int_{0}^{t}\varepsilon\left(
\tau\right)  d\tau\right)  =\frac{C_{0}}{\left(  1+t\right)  ^{a}}+\int
_{0}^{t}\frac{\lambda\left(  t-\xi\right)  d\xi}{\left(  1+\xi\right)  ^{a}}.
\label{W5E6}%
\end{equation}

We now use the methods in Subsection \ref{SlowEps} to approximate
$\lambda\left(  t\right)  $. We will obtain an asymptotics with the form
$\varepsilon\left(  t\right)  \sim\frac{A}{t},$ something that is not
surprising, because the form of (\ref{W5E6}) suggests the behavior $\int_{0}%
^{t}\varepsilon\left(  \tau\right)  d\tau\sim K\log\left(  t\right)  .$ Since
$a\in\left(  0,1\right)  $ we cannot use the method yielding (\ref{W5E2}) but
instead the method yielding (\ref{W6E2a}), (\ref{W6E2}). Suppose that
$\varepsilon\left(  t\right)  \sim\frac{A}{t}$ (something that we will check
``a posteriori"). Then, using (\ref{W6E2a}), (\ref{W6E2}) we obtain
\begin{equation}
\lambda\left(  t\right)  \sim\frac{C}{t^{1+a}}\ \ \text{as\ \ }t\rightarrow
\infty. \label{W5E8a}%
\end{equation}

On the other hand, we can rewrite (\ref{W5E6}) as
\begin{equation}
\frac{d}{dt}\left(  \exp\left(  \int_{0}^{t}\varepsilon\left(  \tau\right)
d\tau\right)  \right)  =\frac{C_{0}}{\left(  1+t\right)  ^{a}}+\int_{0}%
^{t}\frac{\lambda\left(  t-\xi\right)  d\xi}{\left(  1+\xi\right)  ^{a}}.
\label{W5E8}%
\end{equation}
Combining (\ref{W5E8a}) and (\ref{W5E8}) we obtain the approximation
\[
\frac{d}{dt}\left(  \exp\left(  \int_{0}^{t}\varepsilon\left(  \tau\right)
d\tau\right)  \right)  \sim\frac{K}{t^{a}}\ \ \text{as\ \ }t\rightarrow\infty
\]
whence $\int_{0}^{t}\varepsilon\left(  \tau\right)  d\tau\sim\log\left(
t^{1-a}\right)  $ as $t\rightarrow\infty.$ Then $\varepsilon\left(  t\right)
\sim\frac{\left(  1-a\right)  }{t}$ as $t\rightarrow\infty.$ We then recover
the ansatz made for $\varepsilon\left(  t\right)  $ with $A=\left(
1-a\right)  .$

\bigskip

\subsubsection{Conclusions on the toy model}

The model \eqref{A1E4}-\eqref{A1E6} either with $\varepsilon$ constant or
given by \eqref{A1E7} gives some information about the 
particle distribution which evolves according to the combined effect of particle transport
and rare collisions. It is interesting to remark that even in the case of
constant collision rate $\varepsilon$, different moments of the function $f$
are asymptotically given by different functions in a way that is not
compatible with a self- similar behavior for the distribution $f$. 
In the case of the model \eqref{A1E4}-\eqref{A1E6} with $\varepsilon$ given by
\eqref{A1E7} we formally obtained an asymptotic formula for the long time
behavior of the function $\varepsilon(t)$. More precisely, we show that  $\varepsilon\left(  t\right)
\sim\frac{\left(  1-a\right)  }{t}$ as $t\rightarrow\infty.$ Notice that the lack of integrability of this rate as $t\to \infty$ implies that a given particle experiences infinitely many collisions as $t\to\infty$. It would be relevant to prove
rigorously this asymptotic behavior and to extend these results to the full
nonlinear Boltzmann equation.


\subsection{A remark on the homoenergetic solutions for the Fokker-Planck operator}

The problem of some  particular homoenergetic solutions for a different kinetic equation, namely the Fokker-Planck equation has been considered in \cite{MT}. More precisely, instead of considering the Boltzmann equation \eqref{A0_0} they consider 
\begin{align}
\partial_{t}f+v\partial_{x}f  &  
=\Delta_{v} f+\frac{1}{\varepsilon(x)}\partial_{v}\left(f(v-V(x))\right) , \ \label{eq:FP}%
\end{align}
with $V(x)$ and $\varepsilon(x)$ as in \eqref{eq:averagevel} and \eqref{eq:intenergy} respectively. In \cite{MT} the case of simple shear and higher dimensional generalizations of it are considered. Notice that this case, from a dimensional analysis point of view, corresponds to the hyperbolic-dominated case considered in this paper. The following solution of \eqref{eq:FP} is obtained and its stability properties are described. 
\begin{equation}
f\left(  t,x,v\right) =(\det \eta(t))G\big(p\big)\label{eq:MT}
\end{equation}
with 
$$p=\eta(t) \big( v+Kx_2 (1,0)^T\big),$$
\begin{equation}
\eta(t)=\frac{1}{Kt^{\frac{3}{2}}}
\left(
\begin{array}[c]{cc}
\sqrt{3} & 3\\
3 & 2Kt 
\end{array}
\right)  
\end{equation}
and
$$G\big(p)=(4\pi)^{-\frac 1 2 } \exp{\big({-\frac 1 4}\vert p\vert ^2\big)}.$$ 

It is interesting to remark that the solution \eqref{eq:MT} is a stretched Maxwellian. This is very different from the type of behaviour that we obtained for the simplified model introduced in Section \ref{sec:domhyper}. Note also that the models \eqref{A0_0}  and \eqref{eq:FP} are very different from the physical point of view in the hyperbolic-dominated regime. Indeed, in the Boltzmann case the mean free flight time is much larger than the characteristic time in which important shear takes place. On the contrary, the Fokker-Planck dynamics implicitly assumes that the mean free flight time is very small. 

\bigskip

\section{Entropy formulas}
\label{sec:entropy}

\bigskip

Homoenergetic solutions are characterized by constant values in space of the
particle density $\rho=\rho\left(  t\right)  $ and internal energy
$\varepsilon=\varepsilon\left(  t\right)  .$ We are now interested in the form
of another relevant thermodynamic magnitude, namely the entropy that, for the
Boltzmann equation, we identify with minus the $H-$function.

Let be $f=f\left(  t,x,v\right) $ the velocity distribution. We obtain the
following entropy density for particle at a given point $x$
\[
\frac{s\left(  t,x\right)  }{\rho\left(  t\right)  }=-\frac{1}{\rho\left(
t\right)  }\int f\left(  t,x,v\right)  \log\left(  f\left(  t,x,v\right)\right)  d^{3}v.
\]

Using (\ref{B1_0}) it follows that the entropy density for particle is
independent of $x$ and it is given by
\begin{equation}
\frac{s\left(  t\right)  }{\rho\left(  t\right)  }=-\frac{1}{\rho\left(
t\right)  }\int_{\mathbb{R}^{3}}g\left(  t,w\right)  \log\left(  g\left(
t,w\right)  \right)  d^{3}w . \label{U1E8}%
\end{equation}

In a previous paper (cf.~\cite{JNV1}, Section 7) we showed that in the case
of self-similar solutions the formulas for entropy for particle have some
analogies with the corresponding formulas for equilibrium distributions, in
spite of the fact that the distributions obtained there deal with
nonequilibrium situations. This was due to the fact that to a large extent
the entropy formulas depend on the scaling properties of the distributions.

\bigskip

Moreover, there is a case in which the analogy between the entropy formulas
for the equilibrium case and the considered solutions is the largest which
corresponds, nonsurprisingly, to the case in which the particle distribution
is given by Hilbert expansions 
 (cf.~\cite{JNV2}). Indeed, we notice that both in the cases of
solutions given by  time-dependent Maxwellian distributions 
or self-similar solutions we can
approximate $g\left(  t,w\right)  $ as
\begin{equation}
g\left( t, w\right)  \sim\frac{1}{a\left(  t\right)  }G\left(  \frac
{w}{\lambda\left(  t\right)  }\right)  \text{ as }t\rightarrow\infty
\label{U1E7}%
\end{equation}
for suitable functions $a(t), \lambda( t)  $ which
are related to the particle density and the average energy of the particles.

In the case of solutions given by Hilbert expansions the distribution $G$ is a
Maxwellian, which can be assumed to be normalized to have density one and
temperature one. Moreover, we will assume also that the mass of the particles
is normalized to $m=2$ in order to get simpler formulas. This implies that the
Maxwellian distribution takes the form $G_{M}(\xi)=\frac{e^{-|\xi|^{2}}}%
{\pi^{\frac{3}{2}}}.$

We recall that in \cite{JNV1} it has been obtained that
\begin{equation}
\frac{s}{\rho}=\log\left(  \frac{e^{\frac{3}{2}}}{\rho}\right)  +C_{G}
\label{U1E9}%
\end{equation}
where $C_{G}$ is
\begin{equation}
C_{G}=-\frac{\int G\log\left(  G\right)  d\xi}{\int G\left(  \xi\right)  d\xi
}-\log\left[  \frac{\left(  \int\left\vert \xi\right\vert ^{2}Gd\xi\right)
^{\frac{3}{2}}}{\left(  \int G\left(  \xi\right)  d\xi\right)  ^{\frac{5}{2}}
}\right], \label{U2E1}
\end{equation}
and $\frac{s}{\rho}\to\infty$ as $t\to\infty$. 
The formula (\ref{U1E9}) has the same form as the usual formula of the entropy
for ideal gases, except for the value of the constant $C_{G}.$ In the case of
solutions given by Hilbert expansions the value of $C_{G}$ is the same as the
one in the formula of the entropy for ideal gases. Therefore, in the case of
the solutions obtained in  \cite{JNV2} which can be approximated by Hilbert
expansions, the asymptotic formula for the entropy by particle is the same as
the one for ideal gases.

In the case of solutions corresponding to a hyperbolic-dominated behavior the
formula of the entropy does not necessarily resemble the formula of the
entropy for ideal gases, because in general the scaling properties of the
particle distributions are very different from the ones taking place in the
case of gases described by Maxwellian distributions. 

For instance, if the
homoenergetic flow is a homogeneous dilatation (cf. Subsection
\ref{ss:3ddilslow}) the formula (\ref{U1E9}) holds with a constant $C_{G}$ which depends 
on the initial particle distribution $G_{0}.$ However, in this case, $\frac{s}{\rho}$ converges to a finite limit as $t\to\infty$ and therefore the contribution of the constant $C_{G}$ in \eqref{U1E9} is of the same order of magnitude as each of the other two terms. 

In the case of simple shear with $\gamma<-1$ (cf.~Subsection \ref{SimpShearFrColl}) we obtain the following formulas for large
$t$ (cf. (\ref{A3E1})-(\ref{A3E3}))%
\begin{align}
\rho &  =\int G_{0}\left(  \eta_{1},w_{2},w_{3}\right)  d\eta_{1}dw_{2}%
dw_{3}\nonumber\\
\varepsilon &  \sim K^{2}t^{2}\int G_{0}\left(  \eta_{1},w_{2},w_{3}\right)
\left(  w_{2}\right)  ^{2}d\eta_{1}dw_{2}dw_{3}\nonumber\\
s  &  =-\int_{\mathbb{R}^{3}}G_{0}\left(  \eta_{1},w_{2},w_{3}\right)
\log\left(  G_{0}\left(  \eta_{1},w_{2},w_{3}\right)  \right)  d\eta_{1}%
dw_{2}dw_{3}. \label{U2E2}%
\end{align}

Notice that (\ref{U2E2}) implies that the entropy of the distribution does not
increase as $t\rightarrow\infty,$ somehing that it is not surprising given
that the role of the collisions is negligible. The average energy of the
molecules increases due to the shear, but the entropy does not increase. Therefore, (\ref{U1E9}) holds with a constant $C_{G}$ depending on the solution itself but, since the three terms in the equation are of the same order of magnitude as in the case of homogeneous dilatation, the formula does not give much information.

\bigskip In the case of cylindrical dilatation the asymptotic behavior of $G$
is described by (\ref{D3E1}), i.e.
\[
G\left(  \tau,w\right)  =e^{2\tau}G_{\infty}\left(  e^{\tau}w_{1},e^{\tau
}w_{2},w_{3}\right)  .
\]
Then $s$ is given by
\begin{align*}
s  &  =-\int_{\mathbb{R}^{3}}e^{2\tau}G_{\infty}\left(  e^{\tau}w_{1},e^{\tau
}w_{2},w_{3}\right)  \log G_{\infty}\left(  e^{\tau}w_{1},e^{\tau}w_{2}%
,w_{3}\right)  dw\\
&  =-\int_{\mathbb{R}^{3}}G_{\infty}\left(  \eta_{1},\eta_{2},w_{3}\right)
\log G_{\infty}\left(  \eta_{1},\eta_{2},w_{3}\right)  d\eta_{1}d\eta
_{2}dw_{3}.
\end{align*}
We then obtain the same situation as in the case of planar shear (with $K=0$),
i.e.~the entropy does not increase much for large values of $t$ and its limit
value is much dependent on the initial particle distribution.

\bigskip In the case of combined orthogonal shears (cf.~Subsection
\ref{CombShear}) we do not have much information about the distribution of
particles, but we have argued that this distribution cannot be expected to be
self-similar, and very likely the particle distributions and energy
distributions are concentrated in different scales of $\left\vert w\right\vert
.$ Therefore, it is unlikely that formulas like (\ref{U1E9})
could be satisfied in such cases.

\section{Concluding remarks}
\label{sec:tableconcl}

We have described in \cite{JNV1}, \cite{JNV2} and in this paper  
several examples of long time asymptotics for homoenergetic solutions of the
Boltzmann equation. This particular class of solutions exhibits a rich variety of possible asymptotic behaviors. 

As discussed in \cite{JNV2} the key feature which distinguishes the different asymptotic behaviours is the relative size of the collision term and the hyperbolic term in the equation satisfied by the homoenergetic flows (cf.~\eqref{D1_0}). 
The situation in which
the collision terms are the dominant ones for large times has been studied in \cite{JNV2}. The case in which there is a balance between hyperbolic and collision terms has been rigorously analyzed in \cite{JNV1}. In this paper we considered the case in which the hyperbolic
terms are the dominant ones as $t\rightarrow\infty$.

When the hyperbolic terms are much larger than the collision terms the
resulting solutions yield much more complex behaviors than the ones that we
have obtained in the previous cases. One of the reasons for this is that in
some cases the description of the asymptotic behavior of the solutions is a
singular perturbation problem, in which the collision term is very small but
plays a crucial role determining the behavior of the solutions for large
times, because the collisions, in spite of their smallness, yield huge
modifications of the geometry of the velocity distributions. In other cases we
have found that the collisions are so small that their effect becomes
irrelevant as $t\rightarrow\infty.$ These are situations in which the
collision rate becomes so small that the expected number of collisions for a
given particle is bounded as $t\rightarrow\infty.$ In these cases we say that
we have ``frozen collisions".

As it might be seen in this paper,
in the hyperbolic-dominated case, the detailed understanding of the
particle distributions for long times is largely open and challenging. The analysis of these flows suggest many new interesting mathematical questions which deserve further investigation.

\bigskip

\textbf{Acknowledgements. }  We thank Stefan M\"uller, who motivated us to study this problem, for
useful discussions and suggestions on the topic.  
The work of R.D.J. was supported by ONR (N00014-14-1-0714), AFOSR (FA9550-15-1-0207), NSF (DMREF-1629026), and the MURI program (FA9550-18-1-0095, FA9550-16-1-0566). 
A.N. and J.J.L.V. acknowledge support through the
CRC 1060 \textit{The mathematics of emergent effects }of the University of
Bonn that is funded through the German Science Foundation (DFG).

\end{document}